\documentstyle{ar}
\begin{document}
\newcommand{\datetime}
           {{\count0=\time \divide\count0 by 60
             \count1=\count0
             \ifnum\count1>12 \advance\count1 by -12 \fi
             \count2=\count0 \multiply\count2 by -60
             \count3=\time \advance\count3 by \count2
             \today\
             \ifnum\count1<10 0\fi \number\count1:
             \ifnum\count3<10 0\fi \number\count3\
             \ifnum\count0<12 a.m.\else p.m.\fi}}
\newcommand{\ee}[1]{\mbox{${} \times 10^{#1}$}}
\newcommand{\eten}[1]{\mbox{$10^{#1}$}}
\newcommand{\jj}[2]{\mbox{$J = #1\rightarrow#2$}}
\newcommand{\jkkjkk}[6]{\mbox{$J_{K_{-1}K_{+1}}
                              = #1_{#2#3}\rightarrow#4_{#5#6}$}}
\newcommand{\fs}[6]{\mbox{$
                               #1_{#2#3}\rightarrow#4_{#5#6}$}}

\newcommand\h {$^h$}
\newcommand\m {$^m$}
\newcommand\s {$^s$}
\newcommand\am{${'}$}
\newcommand\as{${''}$}
\newcommand{\kms}{\mbox{km s$^{-1}$}}
\newcommand\cmv{\mbox{cm$^{-3}$}}
\newcommand\cmc{\mbox{cm$^{-2}$}}
\newcommand\cmdv{\mbox{cm$^{-2}$ (\kms)$^{-1}$}}
\newcommand{\um}{$\mu$m}
\newcommand{\micron}{$\mu$m}
\newcommand{\x}{\mbox{${}\times{}$}}

\newcommand{\iras}{\mbox{\it IRAS}}
\newcommand{\iso}{\mbox{\it ISO}}
\newcommand{\mm}{millimeter}
\newcommand\submm{submillimeter}
\newcommand\fir{far-infrared}
\newcommand\mir{mid-infrared}
\newcommand\nir{near-infrared}
\newcommand{\sfr }{\mbox{$\dot M_{\star}$}}
\newcommand\sed{spectral energy distribution}
\newcommand{\lsun}{\mbox{L$_\odot$}}
\newcommand{\msun}{\mbox{M$_\odot$}}
\newcommand{\ta}{{$T_A^*$}}
\newcommand{\tex}{\mbox{$T_{\rm ex}$}}
\newcommand{\tmb}{\mbox{$T_{\rm mb}$}}
\newcommand{\tr}{\mbox{$T_R$}}
\newcommand{\tk}{\mbox{$T_K$}}
\newcommand{\td}{\mbox{$T_D$}}
\newcommand{\tbol}{\mbox{$T_{bol}$}} 
\newcommand{\dv}{\mbox{$\Delta v$}}
\newcommand{\n}{\mbox{$n$}}
\newcommand{\nbar}{\mbox{$\overline{n}$}}
\newcommand{\nc}{\mbox{$n_c(jk)$}}
\newcommand{\mv}{\mbox{$M_V$}} 
\newcommand{\mc}{\mbox{$M_N$}} 
\newcommand{\mn}{\mbox{$M_n$}} 
\newcommand{\meanl}{\mbox{$\langle l \rangle$}} 
\newcommand{\meanar}{\mbox{$\langle a/b \rangle$}} 
\newcommand{\mean}[1]{\mbox{$\langle#1\rangle$}} 
\newcommand{\opacity}{\mbox{$\kappa(\nu)$}} 
\newcommand{\av}{\mbox{$A_V$}} 
\newcommand{\bperp}{\mbox{$B_{\perp}$}} 
\newcommand{\rinf}{\mbox{$r_{inf}$}} 

\newcommand{\hh}{\mbox{{\rm H}$_2$}}
\newcommand{\form}{H$_2$CO}
\newcommand{\water}{H$_2$O}
\newcommand{\ammonia}{\mbox{{\rm NH}$_3$}}
\newcommand{\coo}{$^{13}$CO}
\newcommand{\cooo}{C$^{18}$O}
\newcommand{\coooo}{C$^{17}$O}
\newcommand{\hcop}{HCO$^+$}
\newcommand{\hcopi}{H$^{13}$CO$^+$}
\newcommand{\dcop}{DCO$^+$}

\input{epsf}
\newcommand{\psfig}[3]
           {{\begin{figure}[tbp]
             \vbox to#1{\epsfxsize=\textwidth\epsfbox{#2.ps}}
             \caption{#3}
             \end{figure}}}
\newcommand{\psscaledfig}[4]
           {{\begin{figure}[tbp]
             \centerline{\vbox to#1{\epsfxsize=#2\epsfbox{#3.ps}}}
             \caption{#4}
             \end{figure}}}


\title{Physical Conditions in Regions of Star Formation}
\markboth{Evans}{Star Formation}

\author{Neal J. Evans II\affiliation{Department of Astronomy,\\
 The University of Texas at Austin}}

\begin{keywords}
star formation, interstellar molecules, molecular clouds
\end{keywords}

\begin{abstract}
The physical conditions in molecular clouds control the nature and rate
of star formation, with consequences for planet formation and galaxy
evolution. The focus of this review is on the conditions that characterize
regions of star formation in our Galaxy. A review of the tools and
tracers for probing physical conditions includes summaries of generally
applicable results. Further discussion distinguishes between the formation of
low-mass stars in relative isolation and formation in a clustered
environment. Evolutionary scenarios and theoretical predictions are
more developed for isolated star formation, and observational tests
are beginning to interact strongly with the theory. Observers have
identified dense cores collapsing to form individual stars or binaries, and
analysis of some of these support theoretical models of collapse. Stars of both
low and high mass form in clustered environments, but massive stars
form almost exclusively in clusters. The theoretical understanding of
such regions is considerably less developed, but observations
are providing the ground rules within which theory must operate.
The most rich and massive star clusters form in massive, dense,
turbulent cores, which provide models for star formation in other
galaxies.

\end{abstract}

\maketitle

\section{INTRODUCTION}   \label{intro}

Long after their parent spiral galaxies have formed, stars continue to form
by repeated condensation from the interstellar medium.
In the process, parts of the interstellar
medium pass through a cool, relatively dense phase with a great deal of
complexity-- molecular clouds. While both the diffuse interstellar medium
and stars can be supported by thermal pressure, most molecular clouds
cannot be thermally supported (Goldreich \& Kwan 1974).
Simple consideration would suggest that
molecular clouds would be a very transient phase in the conversion of
diffuse gas to stars, but in fact they persist much longer than expected.
During this extended life, they produce an intricate physical
and chemical system that provides the substrate for the formation of planets
and life, as well as stars.
Comparison of cloud masses to the total mass of stars that they produce
indicates that most of the matter in a molecular cloud is sterile; stars
form only in a small fraction of the mass of the cloud (Leisawitz et al 1989).

The physical conditions in the bulk of a molecular cloud provide the key to
understanding why molecular clouds form an essentially metastable state along
the
path from diffuse gas to stars. Most of the mass of most
molecular clouds in our Galaxy is contained in regions of modest extinction,
allowing photons from the interstellar radiation field to maintain
sufficient ionization for magnetic fields to resist collapse (McKee 1989); most
of the molecular gas is in fact in a photon-dominated region, or PDR
(Hollenbach \& Tielens 1997). In addition, most molecular gas has supersonic
turbulence (Zuckerman \& Evans 1974).
The persistence of such turbulence over the inferred
lifetimes of clouds in the face of rapid damping mechanisms (Goldreich \& Kwan
1974) suggests constant replenishment,
most likely in a process of self-regulated star formation
(Norman \& Silk 1980, Bertoldi \& McKee 1996), since star formation
is accompanied by energetic outflows, jets, and winds (Bachiller 1996).

For this review, the focus will be on
the physical conditions in regions that {\it are} forming stars and
likely precursors of such regions.
While gravitational collapse explains the formation of
stars, the details of how it happens depend critically on the physical
conditions in the star-forming region. The details determine the
mass of the resulting star and the amount of mass that winds up in
a disk around the star, in turn controlling
the possibilities for planet formation. The physical conditions also
control the chemical conditions. With the recognition that much
interstellar chemistry is preserved in comets (Crovisier 1999, van Dishoeck \&
Blake 1998), and that interstellar chemistry may also affect planet formation
and the possibilities for life (e.g. Pendleton 1997, Pendleton \& Tielens 1997,
Chyba \& Sagan 1992), the knowledge of physical conditions in star-forming
regions has taken on additional significance.

Thinking more globally, different
physical conditions in different regions determine whether a few,
lightly clustered stars form (the isolated mode) or a tight grouping
of stars form (the clustered mode) (Lada 1992; Lada et al 1993).
The star formation rates per unit mass of molecular gas vary by a factor
$> \eten2$ in clouds within our own Galaxy (Evans 1991, Mead et al 1990),
and starburst galaxies achieve even higher rates than are seen anywhere in our
Galaxy (e.g. Sanders et al 1991). Ultimately, a description of
galaxy formation must incorporate an understanding of how
star formation depends on physical conditions, gleaned from
careful study of our Galaxy and nearby galaxies.

Within the space limitations of this review, it is not possible
to address all the issues raised by the preceding overview.
I will generally avoid topics that have been recently reviewed,
such as circumstellar disks (Sargent 1996, Papaloizou \& Lin 1995,
Lin \& Papaloizou 1996, Bodenheimer 1995), as well as
bipolar outflows (Bachiller 1996), and dense PDRs (Hollenbach \& Tielens 1997).
I will also discuss the sterile parts of molecular clouds only as relevant
to the process that leads some parts of the cloud to be suitable
for star formation.
While the chemistry and physics of star-forming regions
are coupled, chemistry has been recently reviewed (van Dishoeck \& Blake 1998).
Astronomical masers are being concurrently reviewed (Menten 1999); as with
HII regions, they will be discussed only as signposts for regions of
star formation.

I will focus on star formation in our Galaxy.
Nearby regions of isolated, low-mass star formation will receive considerable
attention (\S \ref{isolated})
because we have made the most progress in studying them.
Their conditions will be compared to those in regions forming clusters
of stars, including massive stars (\S \ref{clustered}).
These regions of clustered star formation are poorly understood,
but they probably form the majority of stars in our Galaxy (Elmegreen 1985),
and they are the regions relevant for comparisons to other galaxies.

Even with such a restricted topic, the literature is vast. I make no attempt
at completeness in referencing. On relatively non-controversial topics,
I will tend to give an early reference and a recent review; for more
unsettled topics, more references, with different points of view, will be
given.
Recent or upcoming publications with significant overlap include Hartmann
(1998), Lada \& Kylafis (1999), and Mannings et al (2000).

\section{PHYSICAL CONDITIONS }  \label{defs}

The motivation for studying physical conditions can be found in a few
simple theoretical considerations. Our goal is to know when and
how molecular gas collapses to form stars. In the simplest situation,
a cloud with only thermal support, collapse should occur if the mass
exceeds the Jeans (1928) mass,

\begin{equation}
 M_J = \biggl({\pi k \tk\over\mu m_H G}\biggr)^{1.5} \rho^{-0.5}
 = 18 \msun \tk^{1.5} n^{-0.5},
\end{equation}
where \tk\ is the kinetic temperature (kelvins),
$\rho$ is the mass density (gm \cmv),
and $n$ is the total particle density (\cmv).
In a molecular cloud, H nuclei are almost
exclusively in \hh\ molecules, and $ n \simeq n(\hh) + n({\rm He})$.
Then $\rho = \mu_n m_H n$, where $m_H$
is the mass of a hydrogen atom and $\mu_n$ is the mean mass
per particle (2.29 in a fully molecular cloud with 25\% by mass helium).
Discrepancies between coefficients in the equations presented here and
those in other references usually are traceable to a different definition
of $n$.
In the absence of pressure support, collapse will occur in a free-fall
time (Spitzer 1978),

\begin{equation}
 t_{ff} = \biggl({3 \pi\over32 G \rho}\biggr)^{0.5} =
3.4\ee{7} n^{-0.5} {\rm years}.
\end{equation}
If $\tk = 10$ K and $n \geq 50$ \cmv, typical conditions in the {\it sterile}
regions (e.g. Blitz 1993),
$M_J \leq 80 \msun$, and $t_{ff} \leq 5\ee{6}$ years.
Our Galaxy contains about 1--3\ee{9} \msun\ of molecular gas (Bronfman et al
1988, Clemens et al 1988, Combes 1991).
The majority of this gas is probably
contained in clouds with $M > 10^4$ \msun\ (Elmegreen 1985).
It would be highly unstable on these grounds, and free-fall collapse
would lead to  a star formation rate, $\sfr \geq 200 $ \msun yr$^{-1}$,
far in excess of the recent Galactic average of 3 \msun yr$^{-1}$
(Scalo 1986). This argument, first made by Zuckerman \& Palmer (1974),
shows that most clouds cannot be collapsing at free fall (see also
Zuckerman \& Evans 1974).
Together with evidence of cloud lifetimes of about
$4\ee{7}$ yr (Bash et al 1977; Leisawitz et al
1989), this discrepancy motivates an examination of other support mechanisms.

Two possibilities have been considered, magnetic fields and turbulence.
Calculations of the stability of magnetized clouds (Mestel \&
Spitzer 1956, Mestel 1965) led to the concept of a magnetic critical
mass ($M_B$). For highly flattened clouds,

\begin{equation}
M_B = (2\pi)^{-1} G^{-0.5} \Phi,
\end{equation}
(Li \& Shu 1996), where $\Phi$ is the magnetic flux threading the cloud,

\begin{equation}
\Phi \equiv \int B da.
\end{equation}
Numerical calculations (Mouschovias \& Spitzer 1976) indicate a similar
coefficient (0.13).
If turbulence can be thought of as causing pressure, it may be able to
stabilize clouds on large scales (e.g. Bonazzola et al 1987).
It is not at all clear that turbulence can be treated
so simply. In both cases, the cloud can only be metastable. Gas can move
along field lines and ambipolar diffusion will allow neutral gas to
move across field lines with a timescale of
(McKee et al 1993)

\begin{equation}
t_{AD} = {3\over4 \pi G \rho \tau_{ni}} \simeq 7.3\ee{13} x_e {\rm years},
\end{equation}
where $\tau_{ni}$ is the ion-neutral collision time. The ionization fraction
($x_e$) depends on ionization by photons and cosmic rays, balanced by
recombination.
It thus depends on the abundances of other species ($X(x) \equiv n(x)/n$).

These two
suggested mechanisms of cloud support (magnetic and turbulent) are not entirely
compatible because turbulence should tangle the magnetic field (compare the
reviews by Mouschovias 1999 and McKee 1999).
A happy marriage between magnetic fields and turbulence
was long hoped for; Arons and Max (1975) suggested that
magnetic fields would slow the decay of turbulence if the turbulence was
sub-Alfv\'enic. Simulations of MHD turbulence in systems with high degrees
of symmetry supported this suggestion and indicated that the pressure
from magnetic
waves could stabilize clouds (Gammie \& Ostriker 1996).
However, more recent 3-D simulations indicate that MHD turbulence
decays rapidly and that replenishment is still needed (Mac Low et al 1998,
Stone et al 1999). The usual suggestion is that outflows generate turbulence,
but Zweibel (1998) has suggested that an instability induced by ambipolar
diffusion may convert magnetic energy into turbulence. Finally, the issue
of supporting clouds assumes a certain stability and cloud integrity
that may be misleading in a dynamic interstellar medium
(e.g. Ballesteros-Paredes et al 1999). For a current review of this
field, see V\'azquez-Semadini et al (2000).

With this brief and simplistic review of the issues of cloud stability
and evolution, we have motivated the study of the basic physical conditions:

\begin{equation}
\tk, n, \vec v, \vec B, X.
\end{equation}
All these are local variables; in principle, they can have different values
at each point ($\vec r$) in space and can vary with time ($t$). In practice,
we usually can measure only one component of vector quantities, integrated
through the cloud. For example, we measure only line-of-sight velocities,
usually characterized by the linewidth ($\Delta v$) or higher moments,
and the line-of-sight magnetic field ($B_z$) through the
Zeeman effect, or the projected direction (but not the strength)
of the field in the plane of the sky (\bperp) by polarization studies.
In addition, our observations always
average over finite regions, so we attempt to simplify the dependence on
$\vec r$ by assumptions or models of cloud structure. In \S \ref{probes},
I will describe the methods used to probe these quantities
and some overall results.
Abundances have been reviewed recently (van Dishoeck \& Blake 1998),
so only relevant results will be
mentioned, most notably the ionization fraction, $x_e$.

In addition to the local variables, quantities that explicitly integrate
over one or more dimensions are often measured. Foremost is the column density,

\begin{equation}
N \equiv \int n dl.
\end{equation}
The extinction or optical depth of dust at some wavelength is a common
surrogate
measure of $N$. If the column density is integrated over an area,
one measure of the cloud mass within that area is obtained:

\begin{equation}
\mc \equiv \int N da.
\end{equation}
Another commonly used measure of mass is obtained by simplification of the
virial theorem. If external pressure and magnetic fields are ignored,

\begin{equation}
\mv = C_v G^{-1} R \Delta v^2 = 210 \msun C_v R(pc) (\Delta v(\kms))^2,
\end{equation}
where $R$ is the radius of the region, $\Delta v$ is the FWHM linewidth, and
the constant ($C_v$) depends on geometry and cloud structure, but is of order
unity (e.g. McKee \& Zweibel 1992, Bertoldi \& McKee 1992).
A third mass estimate can be obtained by integrating the density over the
volume,

\begin{equation}
\mn \equiv \int n dv.
\end{equation}
\mn\ is commonly used to estimate $f_v$, the volume filling factor of gas
at density $n$, by dividing \mn\ by another mass estimate, typically \mv.
Of the three methods of mass determination, the virial mass is the least
sensitive to uncertainties in distance and size, but care must be taken
to exclude unbound motions, such as outflows.

Parallel to the physical conditions in the gas, the dust can be characterized
by
a set of conditions:

\begin{equation}
\td, n_D, \opacity,
\end{equation}
where $T_D$ is the dust temperature,
$n_D$ is the density of dust grains,
and \opacity\ is the opacity at a given frequency, $\nu$. If we look in more
detail, grains have a range of sizes (Mathis et al 1977) and compositions.
For smaller grains, \td\ is a function of grain size.
Thus, we would have to characterize the temperature distribution as a
function of size,
the composition of grains, both core and mantle, and many more optical
constants
to capture the full range of grain properties.
For our purposes, \td\ and \opacity\ are the most important properties,
because \td\ affects gas energetics and \td\ and \opacity\ control
the observed continuum emission of molecular clouds.
The detailed nature of the dust
grains may come into play in several situations, but the primary observational
manifestation of the dust is its ability to absorb and emit radiation. For this
review,
a host of details can be ignored. The optical depth is set by

\begin{equation}
\tau_D(\nu) = \opacity N,
\end{equation}
where it is convenient to define \opacity\ so that $N$ is the gas
column density rather than the dust column density.
Away from resonances, the opacity is usually approximated by
$\opacity \propto \nu^{\beta}$.

\section{PROBES OF PHYSICAL CONDITIONS} \label{probes}

The most fundamental fact about molecular clouds is that most of their
contents are invisible. Neither the \hh\ nor the He in the bulk of the clouds
are excited sufficiently to emit. While fluorescent emission from \hh\
can be mapped over the face of clouds (Luhman \& Jaffe 1996), the ultraviolet
radiation
needed to excite this emission does not penetrate the bulk of the cloud.
In shocked regions, \hh\ emits rovibrational lines that are useful probes
of \tk\ and $\vec v(\vec r)$ (e.g. Draine \& McKee 1993).
Absorption by \hh\ of background stars is also difficult: the dust in
molecular clouds obnubilates the ultraviolet that would reveal electronic
transitions; the rotational transitions are so weak that only huge $N$
would produce absorption, and the dust again obnubilates background sources;
only the vibrational transitions in the \nir\ have been seen
in absorption in only a few molecular clouds (Lacy et al 1994).
Gamma rays resulting from cosmic ray interactions with atomic nuclei do probe
all the material in molecular clouds (Bloemen 1989, Strong et al
1994).  So far, gamma-ray studies have suffered from low spatial resolution
and uncertainties in the cosmic ray flux; they have been used mostly to check
consistency with other tracers on large scales.

In the following subsections,
I will discuss probes of different physical quantities, including some general
results, concluding with a discussion of the observational and analytical
tools.  Genzel (1992) has presented a detailed discussion of probes of
physical conditions.

\subsection{Tracers of Column Density, Size, and Mass}  \label{probecol}

Given the reticence of the bulk of the \hh, essentially all probes of
physical conditions rely on trace constituents, such as dust particles
and molecules other than \hh. Dust particles
(Mathis 1990, Pendleton \& Tielens 1997) attenuate light
at short wavelengths (ultraviolet to \nir) and emit at longer
wavelengths (\fir\ to \mm). Assuming that the ratio of dust extinction at
a fixed wavelength to gas column density is constant,
one can use extinction
to map $N$ in molecular clouds, and early work at visible wavelengths
revealed locations and sizes of many molecular clouds before they were
known to contain molecules (Barnard 1927, Bok \& Reilly 1947, Lynds 1962).
There have been more recent surveys for small clouds (Clemens \& Barvainis
1988) and for clouds at high latitude (Blitz et al 1984).
More recently, \nir\ surveys have been
used to probe much more deeply; in particular, the $H-K$ color excess
can trace $N$ to an equivalent visual extinction, $A_V \sim 30$ mag
(Lada et al 1994, Alves et al 1998a). This method provides many pencil
beam measurements through a cloud toward background stars. The
very high resolution, but very undersampled, data require careful analysis
but can reveal information on mass, large-scale structure in $N$ and
unresolved structure (Alves et al 1998a).
Padoan et al (1997) interpret the data of Lada et al (1994) in terms
of a lognormal distribution of density, but Lada et al (1999) show that
a cylinder with a density gradient, $n(r) \propto r^{-2}$, also matches
the observations.

Continuum emission from dust at long wavelengths is complementary to
absorption studies (e.g. Chandler \& Sargent 1997). Because the dust
opacity decreases with increasing wavelength ($\opacity \propto \nu^{\beta}$,
with $\beta \sim 1-2$), emission at long wavelengths can trace large column
densities and provide independent mass estimates (Hildebrand 1983).
The data can be fully sampled and have reasonably high resolution.
The dust emission depends on the dust temperature (\td), linearly
if the observations are firmly in the Rayleigh-Jeans limit,
but exponentially on the Wien side of the blackbody curve. Observations
on both sides of the emission peak can constrain \td.
Opacities have been calculated for a variety of scenarios including
grain mantle formation and collisional concrescence (e.g. Ossenkopf \&
Henning 1994). For grain sizes much less than
the wavelength, $\beta \sim 2$ is expected from simple grain models,
but observations of dense regions often indicate lower values.
By observing at a sufficiently long wavelength, one can trace
$N$ to very high values. Recent results indicate that
$\tau_D(1.2 mm) = 1$ only for  $A_V \sim 4\ee4$ mag (Kramer et al 1998a) in
the less dense regions of molecular clouds.
In dense regions, there is considerable evidence for increased grain opacity
at long wavelengths (Zhou et al 1990, van der Tak et al 1999), suggesting
grain growth through collisional concrescence
in addition to the formation of icy mantles. Further growth of grains in
disks is also likely (Chandler \& Sargent 1997).

The other choice is to use a trace constituent of the gas, typically
molecules that emit in their rotational transitions at \mm\ or \submm\
wavelengths. By using the appropriate transitions of the appropriate
molecule, one can tune the probe to study the physical quantity of interest
and the target region along the line of sight. This technique was first
used with OH (Barrett et al 1964), and it has
been pursued intensively in the 30 years since the discovery of polyatomic
interstellar molecules (Cheung et al 1968; van Dishoeck \& Blake 1998).

The most abundant molecule after \hh\ is carbon monoxide; the main isotopomer
($^{12}$C$^{16}$O) is usually written simply as CO. It is the most common
tracer of molecular gas.  On the largest scales,
CO correlates well with the gamma-ray data (Strong et al 1994), suggesting
that the overall mass of a cloud can be measured even when the line is quite
opaque. This stroke of good fortune can be understood if the clouds
are clumpy and macroturbulent, with little radiative coupling between clumps
(Wolfire et al 1993); in this case, the CO luminosity is proportional to
the number of clumps, hence total mass.
Most of the mass estimates for the larger clouds and for the total molecular
mass
in the Galaxy and in other galaxies are in fact based on CO.
On smaller scales, and in regions of high column density, CO fails to trace
column density, and progressively rarer isotopomers are used to trace
progressively higher values of $N$.
Dickman (1978) established a strong correlation of visual
extinction $A_V$ with \coo\ emission for $1.5 \leq A_V \leq 5$. Subsequent
studies have used \cooo\ and \coooo\ to trace still higher $N$
(Frerking et al 1982).
These rarer isotopomers will not trace the outer parts of the cloud,
where photodissociation affects them more strongly than the common ones,
but we are concerned with the more opaque regions in this review.

When comparing $N$ measured by dust emission with $N$ traced by CO
isotopomers, it is important to correct for the fact that emission from low-$J$
transitions of optically thin isotopomers of CO decreases with \tk,
while dust emission increases with \td\ (Jaffe et al 1984).
Observations of many transitions can avoid this
problem but are rarely done. To convert $N({\rm CO})$ to $N$ requires knowledge
of the abundance, $X({\rm CO})$. The only direct measure gave $X({\rm CO})
= 2.7\ee{-4}$ (Lacy et al 1994),
three times greater than inferred from indirect means
(e.g. Frerking et al 1982).
Clearly, this area needs increased attention,
but at least a factor of three uncertainty must be admitted.
Studies of some particularly opaque regions in molecular clouds indicate
severe depletion (Kuiper et al 1996, Bergin \& Langer 1997), raising the
concern that even the rare CO isotopomers may fail to trace $N$.
Indeed, Alves et al (1999) find that \cooo\ fails to trace column
density above $\av = 10$ in some regions, and Kramer et al (1999) argue
that this failure is best explained by depletion of \cooo.

Sizes of clouds, characterized by either a radius ($R$) or diameter ($l$),
are measured by mapping the cloud in a particular tracer; for non-spherical
clouds, these are often the geometric mean of two dimensions, and the
aspect ratio ($a/b$) characterizes the ratio of long and short axes.
The size along the line of sight (depth) can only be constrained by making
geometrical assumptions (usually of a spherical or cylindrical cloud).
One possible probe of the depth is H$_3$$^+$, which is unusual in having
a calculable, constant density in molecular clouds.
Thus, a measurement of $N({\rm H}_3^+)$ can yield
a measure of cloud depth (Geballe \& Oka 1996).

With a measure of size and a measure of column density, the mass (\mc) may
be estimated (equation 8); with a size and a linewidth (\dv),
the virial mass (\mv) can be estimated (equation 9).
On the largest scales, the mass is often estimated from integrating the CO
emission over the cloud and using an empirical relation between mass and the
CO luminosity, $L({\rm CO})$. The mass distribution has been estimated for
both clouds and clumps within clouds, primarily from CO, \coo, or \cooo,
using a variety of techniques to define clumps and estimate masses (Blitz 1993,
Kramer et al 1998b, Heyer \& Terebey 1998, Heithausen et al 1998).
These studies have covered a wide range of masses, with Kramer et al
extending the range down to $M = \eten{-4} \msun$.
The result is fairly well agreed on:
$dN(M) \propto M^{-\alpha} dM$,
with $1.5 \leq \alpha \leq 1.9$.
Elmegreen \& Falgarone (1996) have
argued that the mass spectrum is a result of the fractal nature of the
interstellar gas, with a fractal dimension, $D = 2.3\pm0.3$. There is
disagreement over whether clouds are truly fractal or have a preferred
scale (Blitz \& Williams 1997). The latter authors suggest a scale
of 0.25--0.5 pc in Taurus based on \coo. On the other hand, Falgarone
et al (1998), analyzing an extensive data set, find evidence for continued
structure down to 200 AU in gas that is not forming stars.
The initial mass function (IMF) of stars is steeper than the cloud mass
distribution for $M_\star > 1 \msun$ but is flatter than the cloud mass
function for $M_\star < 1 \msun$ (e.g. Scalo 1998). Understanding the origin
of the differences is a major issue
(see Williams et al 2000, Meyer et al 2000).

\subsection{Probes of Temperature and Density}  \label{probetemp}

The abundances of other molecules are so poorly constrained that
CO isotopomers and dust are used almost exclusively to constrain
$N$ and \mc. Of what use are the over 100 other molecules?
While many are of interest only for astrochemistry, some are very
useful probes of physical conditions like $\tk, n, v, B_z,$ and $x_e$.

Density ($n$) and gas temperature (\tk) are both measured by determining
the populations in molecular energy levels and comparing the results to
calculations of molecular excitation. A useful concept is the excitation
temperature (\tex) of a pair of levels, defined to be the temperature that
gives the actual ratio of populations in those levels, when substituted into
the Boltzmann equation.
In general, collisions and radiative
processes compete to establish level populations; when lines are optically
thick, trapping of line photons enhances the effects of collisions. For some
levels in some molecules, radiative rates are unusually low, collisions
dominate, $\tex = \tk$,  and observational determination of these
``thermalized"
level populations yields \tk. Un-thermalized level populations depend on
both $n$ and \tk; with a knowledge of \tk, observational determination of
these populations yields $n$, though trapping usually needs to be accounted
for.  While molecular excitation probes the local $n$ and \tk\ in principle,
the observations themselves always involve some average over the finite
beam and along the line of sight. Consequently, a model of the cloud
is needed to interpret the observations. The simplest model is of course
a homogeneous cloud, and most early work adopted this model, either explicitly
or implicitly.

Tracers of temperature include CO, with its unusually low dipole moment,
and molecules in which transitions between certain levels are forbidden
by selection rules. The latter include  different $K$ ladders
of symmetric tops like \ammonia, CH$_3$CN, etc. (Ho \& Townes 1983,
Loren \& Mundy 1984).
Different $K_{-1}$ ladders in \form\ also probe \tk\ in dense, warm regions
(Mangum and Wootten 1993).
A useful feature of CO is that its low-$J$ transitions are
both opaque and thermalized in most parts of molecular clouds. In this case,
observations of a single line provide the temperature, after correction for
the cosmic background radiation and departures from the Rayleigh-Jeans
approximation (Penzias et al 1972, Evans 1980).
Early work on CO (Dickman 1975) and
\ammonia\ (Martin \& Barrett 1978) established
that $\tk \simeq 10$ K far from regions of star formation
and that sites of massive star formation are marked by elevated
\tk, revealed by peaks in maps of CO (e.g. Blair et al 1975).

The value of \tk\ far from local heating sources
can be understood by balancing cosmic ray heating and molecular cooling
(Goldsmith \& Langer 1978),
while elevated values of \tk\ in star forming regions
have a more intricate explanation. Stellar photons, even when degraded to
the infrared, do not couple well to molecular gas, so the heating goes
via the dust. The dust is heated by photons and the gas is heated by
collisions with the dust (Goldreich \& Kwan 1974); above a density of
about \eten{4} \cmv, \tk\ becomes well coupled to \td\ (Takahashi et al 1983).
Observational comparison of \tk\ to \td, determined from \fir\ observations,
supports this picture (e.g. Evans et al 1977, Wu \& Evans 1989).

In regions
where photons in the range of 6 to 13.6 eV impinge directly on molecular
material, photoelectrons ejected from dust grains can heat the gas
very effectively and \tk\ may exceed \td. These PDRs (Hollenbach
\& Tielens 1997) form the surfaces of all clouds, but the regions affected
by these photons are limited by dust extinction to about $A_V \sim 8$ mag
(McKee 1989). However, the CO lines often do form in the PDR regions, raising
the
question of why they indicate that $\tk \sim 10$K. Wolfire et al (1993) explain
that the optical depth in the lower $J$ levels usually observed reaches unity
at
a place where the \tk\ and $n$ combine to produce an excitation temperature
(\tex) of about 10 K. Thus, the agreement of \tk\ derived from CO with
the predictions of energetics calculations
for cosmic ray heating may be fortuitous. The \tk\ derived from \ammonia\
refer to more opaque regions and are more relevant to cosmic ray heating.
Finally, in localized regions, shocks can heat the gas to very
high \tk; values of 2000 K are observed in \hh\ ro-vibrational emission
lines (Beckwith et al 1978).  It is clear that characterizing clouds by a
single
\tk, which is often done for simplicity, obscures a great deal of complexity.

Density determination requires observations of several transitions that
are {\it not} in local thermodynamic equilibrium (LTE).
Then the ratio of populations, or equivalently \tex,
can be used to constrain density. A useful concept is the
critical density for a transition from level $j$ to level $k$,

\begin{equation}
\nc = A_{jk}/\gamma_{jk},
\end{equation}
where $A_{jk}$ is the Einstein A coefficient and $n\gamma_{jk}$ is the
collisional deexcitation rate per molecule in level $j$. In general,
both \hh\ and He are effective collision partners, with comparable collision
rates, so that excitation techniques measure the total density of collision
partners, $n \simeq n(\hh) + n({\rm He})$. In some regions of high $x_e$,
collisions with
electrons may also be significant. Detection of a particular transition
is often taken to imply that $n \geq \nc$, but this statement is too
simplistic. Lines can be seen over a wide range of $n$, depending on
observational sensitivity, the frequency of the line, and the optical depth
(e.g. Evans 1989).  Observing high frequency transitions,
multilevel excitation effects, and trapping all
tend to lower the effective density needed to detect a line.
Table 1 contains information for some commonly observed lines, including
the frequency, energy in K above the effective ground state ($E_{up}(K)$),
and the critical densities at $\tk = 10$ K and 100 K. For comparison,
the Table also has $n_{eff}$, the density needed to produce a line of 1 K,
easily observable in most cases.
The values of $n_{eff}$ were calculated with a Large Velocity Gradient
(LVG) code (\S \ref{probetools}) to account for trapping,
assuming log$(N/\dv) = 13.5$ for all
species but \ammonia, for which log$(N/\dv) = 15$ was used. $N/\dv$ has units
of \cmdv. These column densities are
typical and produce modest optical depths. Note that $n_{eff}$ can be as
much as a factor of 1000 less than the critical density, especially for
high excitation lines and high \tk.
Clearly, the critical density should be used as a guideline only;
more sophisticated analysis is necessary to infer densities.

Assuming knowledge of \tk, at least two transitions with different $\nc$
are needed to determine both
$n$ and the line optical depth, $\tau_{jk} \propto N_{k}/\dv $, which
determines
the amount of trapping, and more transitions are desirable.
Because $A_{J,J-1} \propto J^3$, where $J$ is the quantum number
for total angular momentum, observing many transitions up a rotational
energy ladder provides a wide range of $\nc$. Linear molecules, like
HCN and \hcop, have been used in this way, but higher levels often
occur at wavelengths with poor atmospheric transmission. Relatively heavy
species, like CS, have many accessible transitions, and up to five
transitions ranging up to $J=10$ have been used to constrain density
(e.g. Carr et al 1995, van der Tak et al 1999). More complex species
provide more accessible energy levels; transitions within a single $K_{-1}$
ladder of \form\ provide a valuable density probe (Mangum \& Wootten 1993).
Transitions of \form\ with $\Delta J = 0$ are accessible to large arrays
operating at centimeter wavelengths (e.g. Evans et al 1987). The lowest
few of these \form\ transitions have the interesting property of absorbing the
cosmic background radiation (Palmer et al 1969),
\tex\ being cooled by collisional pumping (Townes \& Cheung 1969).

Application of these techniques to the homogeneous cloud model generally
produces estimates of density exceeding \eten{4}\ \cmv\ in regions forming
stars, while the sterile regions of the cloud are thought to have typical
$n \sim 10^2 - 10^3$ \cmv, though these are less well constrained.
Theoretical simulations of turbulence have predicted lognormal
(V\'azquez-Semadini 1994) or power-law (Scalo et al 1998) probability
density functions.
Studies of multiple transitions of different molecules with a wide range
of critical densities often reveal evidence for density inhomogeneities;
in particular, pairs of transitions with higher critical densities tend
to indicate higher densities (e.g. Evans 1980, Plume et al 1997).
Both density gradients and clumpy structure
have been invoked to explain these results (see \S \ref{isolated} and \S
\ref{clustered} for detailed discussion). Since lines with high $\nc$
are excited primarily at higher $n$, one can avoid to some extent the
averaging over the line of sight by tuning the probe.

\subsection{Kinematics}    \label{probekin}

In principle, information on $\vec v (\vec r)$ is contained in maps of
the line profile over the cloud. In practice, this message has been difficult
to decode. Only motions along the line of sight produce Doppler shifts,
and the line profiles average over the beam and along the line of sight.
Maps of the line center velocity
generally indicate that the typical cloud is experiencing neither overall
collapse (Zuckerman \& Evans 1974) nor rapid rotation (Arquilla \& Goldsmith
1986, Goodman et al 1993).
Instead, most clouds appear to have velocity fields dominated by
turbulence, because the linewidths
are usually much greater than expected from thermal broadening.
While such turbulence can explain the breadth of the lines, the line
profile is not easily matched.
Even a homogeneous cloud will tend to develop an
excitation gradient in unthermalized lines because of trapping, and
gradients in \tk\ or $n$ toward embedded sources should exacerbate this
tendency.  Simple microturbulent models with decreasing $\tex(r)$
predict that self-reversed line profiles should be seen
more commonly than they are.  Models with many small clumps and macroturbulence
have had some success in avoiding self-reversed line profiles (Martin et al
1984,
Wolfire et al 1993, Falgarone et al 1994, Park \& Hong 1995).

The average linewidths of clouds are larger for larger clouds,
the linewidth-size relation: $\dv \propto R^{\gamma}$ (Larson 1981).
For clouds with the same $N$, the virial theorem would predict
$\gamma = 0.5$, consistent with the results of many studies
of clouds as a whole (e.g. Solomon et al 1987).
Myers (1985) summarized the different relations and distinguished between
those comparing clouds as a whole and those studying trends within a
single cloud.
The status of linewidth-size relations {\it within} clouds, particularly in
star-forming regions, will be discussed in later sections.

\subsection{Magnetic Field and Ionization}  \label{probemag}

The magnetic field strength and direction are important but difficult
to measure. Heiles et al (1993) review the observations and McKee et
al (1993) review theoretical issues.
The only useful measure of the strength is the Zeeman effect,
which probes the line-of-sight field, $B_z$.
Observations of HI can provide some useful probes of $B_z$ in PDRs
(e.g. Brogan et al 1999), but cannot probe the bulk of molecular gas.
Molecules suitable for Zeeman effect measurements have unpaired electrons
and their resulting reactivity tends to decrease their abundance in the denser
regions (e.g. Sternberg et al 1997).
Almost all work has been done with OH, along with some work with CN
(Crutcher et al 1996), but future prospects include
CCS and excited states of OH and CH (Crutcher 1998).
Measurements of $B_z$ have been made with thermal emission or absorption by OH
(e.g. Crutcher et al 1993),
mostly probing regions with $n \sim \eten{3}$ \cmv, where $B_z \simeq 20$
$ \mu$G or with OH maser emission, probing much denser gas, but with less
certain
conditions. As reviewed by Crutcher (1999a), the results for 14 clouds of
widely varying mass with good Zeeman detections indicate that $M_B$ is usually
within a factor of 2 of the cloud mass. Given uncertainties, this result
suggests that clouds with measured $B_z$ lie close to the critical-subcritical
boundary (Shu et al 1999).
The observations can be fit with
$B \propto n^{0.47}$, remarkably consistent with predictions
of ambipolar diffusion calculations (e.g. Fiedler \& Mouschovias 1993).
However, if turbulent motions in clouds are constrained to be comparable
to the Alfv\'en velocity, $v_a \propto B n^{-0.5}$, this result is also
expected (Myers \& Goodman 1988; Bertoldi \& McKee 1992).

The magnetic field direction, projected on the plane of the sky, can be
measured because spinning, aspherical grains tend to align their spin
axes with the magnetic field direction (see Lazarian et al
1997 for a list of mechanisms). Then the dust grains absorb and emit
preferentially in the plane perpendicular to the field. Consequently,
background star light will be preferentially polarized along \bperp\ and
thermal emission from the grains will be polarized perpendicular to \bperp\
(Hildebrand 1988). Goodman (1996) has shown that the grains
that polarize background starlight do not trace the field very deeply into
the cloud, but maps of polarized {\it emission} at \fir, \submm\
(Schleuning 1998) and \mm\ (Akeson \& Carlstrom 1997, Rao et al 1998)
wavelengths are beginning to provide maps of field direction deep into clouds.
Line emission may also be weakly polarized under some conditions (Goldreich
\& Kylafis 1981), providing a potential probe of \bperp\ with velocity
information.  After many attempts, this effect has been detected recently
(Greaves et al 1999).

The ionization fraction ($x_e$) is determined by chemical analysis
and has been discussed by van Dishoeck and Blake (1998). Theoretically,
$x_e$ should drop from about \eten{-4}\ near the outer edge of the cloud to
about \eten{-8}\ in interiors shielded from ultraviolet radiation.
Observational estimates of $x_e$ are converging on values around \eten{-8}\ to
\eten{-7}\ in cores (de Boisanger et al 1996, Caselli et al 1998, Williams
et al 1998, Bergin et al 1999).

\subsection{Observational and Analytical Tools} \label{probetools}

Having discussed how different physical conditions are probed,
I will end this section with a brief summary of the observational
and analytical tools that are used. Clearly, most information
on physical conditions comes from observations of molecular lines.
Most of these lie at \mm\ or \submm\ wavelengths, and progress in this
field has been driven by the development of large single-dish telescopes
operating at \submm\ wavelengths and by arrays of antennas operating
interferometrically at \mm\ wavelengths (Sargent \& Welch 1993).
The \submm\ capability has allowed the study of high-$J$ levels for
excitation analysis and increased sensitivity to dust continuum emission,
which rises with frequency ($S_\nu \propto \nu^2$ or faster).
Studies of \mm\  and \submm\ emission from dust have been greatly
enhanced recently with the development of cameras on single dishes, both
at \mm\ wavelengths (Kreysa 1992) and at \submm\ wavelengths, with
SHARC (Hunter et al 1996) and SCUBA (Cunningham et al 1994).
Examples of the maps that these cameras are producing are the color
plates showing the 1.3 mm emission from the $\rho$ Ophiuchi region
(Motte et al 1998, Figure 1) and the 850 \micron\ and 450 \micron\
emission from the ridge in Orion (L1641) (Johnstone \& Bally 1999, Figure 2).

Interferometric arrays, operating at \mm\ wavelengths,
have provided unprecedented angular resolution (now
better than 1\as ) maps of both molecular line and continuum emission.
They are particularly critical for separating the continuum emission from
a disk and the envelope and for studying deeply embedded binaries
(e.g. Looney et al 1997, Figure 3).
Complementary information has been provided in the infrared,
with \nir\ star-counting (Lada et al 1994, 1999), \nir\ and \mir\ spectroscopy
of rovibrational transitions (e.g. Mitchell et al 1990, Evans et al 1991,
Carr et al 1995, van Dishoeck et al 1998), and \fir\ continuum and
spectral line studies (e.g. Haas et al 1995). Early results from the
{\it Infrared Space Observatory} can be found in Yun \& Liseau (1998).

The analytical tools for molecular cloud studies have grown gradually
in sophistication. Early studies assumed LTE excitation, an approximation
that is still used in some studies of CO isotopomers, but it is clearly
invalid for other species.
Studies of excitation require solution of the statistical equilibrium equations
(Goldsmith 1972). Goldreich \& Kwan (1974) pointed out that photon trapping
will increase the average \tex\ and provided a way of including
its effects that was manageable with the limited computer resources of that
time:
the Large Velocity Gradient (LVG) approximation. Tied originally to their
picture of collapsing clouds, this approximation allowed one to
treat the radiative transport
locally. Long after the overall collapse scenario had been discarded, the LVG
method has remained in use, providing a quick way to include trapping, at least
approximately. In parallel, more computationally intensive codes were developed
for microturbulent clouds, in which photons emitted anywhere in the cloud could
affect excitation anywhere else (e.g. Lucas 1974).

The microturbulent and LVG
assumptions
are the two extremes, and real clouds probably lie between. For modest optical
depths, the conclusions of the two methods differ by factors of about 3,
comparable to uncertainties caused by uncertain geometry (White 1977, Snell
1981).
These methods are still useful in some situations, but they are gradually being
supplanted by more flexible radiative transport codes, using either the
Monte Carlo technique (Bernes 1979, Choi et al 1995, Park \& Hong 1998,
Juvela 1997) or $\Lambda$-iteration (Dickel \& Auer 1994, Yates et al 1997,
Wiesemeyer 1999).
Some of these codes allow variations in the velocity, density, and temperature
fields, non-spherical geometries, clumps, etc. Of course, increased flexibility
means more
free parameters and the need for more extensive observations to constrain them.

Similar developments have occurred in the area of dust continuum emission.
Since stellar photons are primarily at wavelengths where dust is quite opaque,
a radiative transport code is needed to compute dust
temperatures as a function of distance from a stellar heat source (Egan
et al 1988, Manske \& Henning 1998). For clouds without embedded stars
or protostars, only the interstellar radiation field heats the dust;
\td\ can get very low (5--10 K) in centers of opaque clouds (Leung 1975).
Embedded sources heat clouds internally; in clouds opaque to the stellar
radiation, it is absorbed close to the source and reradiated at longer
wavelengths.
Once the energy is carried primarily by photons at wavelengths where the
dust is less opaque,
the temperature distribution relaxes to the optically thin limit (Doty \& Leung
1994):

\begin{equation}
\td(r) \propto L^{q/2} r^{-q}
\end{equation}
where $L$ is the luminosity of the source and $q = {2/(\beta + 4)}$,
assuming $\opacity \propto \nu^{\beta}$.

\section{FORMATION OF ISOLATED LOW-MASS STARS} \label{isolated}

Many low-mass stars actually form in regions of
high-mass star formation, where clustered formation is the rule (Elmegreen
1985,
Lada 1992, McCaughrean \& Stauffer 1994).
The focus here is on regions where we can isolate the individual
star-forming events and these are almost inevitably forming low-mass stars.

\subsection{Theoretical Issues}                \label{theoryissues}

The theory of isolated star formation has been developed in some detail.
It relies on the existence of relatively isolated regions of enhanced
density that can collapse toward a single center, though processes at
smaller scales may cause binaries or multiples to form. One issue then
is whether isolated regions suitable for forming individual, low-mass stars
are clearly identifiable. The ability to separate these from the rest of
the cloud underlies the distinction between sterile and fertile parts of
clouds (\S \ref{intro}).

Because of the enormous compression needed, gravitational collapse
plays a key role in all star formation theories. In most cases, only a part
of the cloud collapses, and theories differ on how this part is
distinguished from the larger cloud.  Is it brought to the verge
of collapse by an impulsive event, like a shock wave (Elmegreen \& Lada
1977) or a collision between clouds or clumps (Loren 1976),
or is the process gradual?  Among gradual
processes, the decay of turbulence and ambipolar diffusion are leading
contenders.  If the decay of turbulence leaves the cloud in a subcritical
state ($M < M_B$), then a relatively long period of ambipolar diffusion
is needed before dynamical collapse can proceed (\S \ref{defs}).
If the cloud is supercritical ($M> M_B$), then the magnetic field alone
cannot stop the collapse (e.g. Mestel 1985).
If turbulence does not prevent it, a rapid collapse ensues, and fragmentation
is likely. Shu et al (1987a, 1987b) suggested that the subcritical case
describes
isolated low mass star formation, while the supercritical case describes
high-mass and clustered star formation. Recently, Nakano (1998) has argued that
star formation in subcritical cores via ambipolar diffusion is implausible;
instead he favors dissipation of turbulence as the controlling factor.
For the present section, the questions are
whether there is evidence that isolated, low-mass stars form in
sub-critical regions and what is the status of turbulence in these cores.

Rotation could in principle support clouds against collapse, except along
the rotation axis (Field 1978).
Even if rotation does not prevent the collapse, it
is likely to be amplified during collapse, leading at some point to
rotation speeds able to affect the collapse. In particular, rotation
is usually invoked to produce binaries or multiple systems on small
scales. What do we know about rotation rates on large scales and how
the rotation is amplified during collapse? Is there any correlation
between rotation and the formation of binaries observationally?
If rotation controls whether binaries form, can we understand why
collapse leads to binary formation roughly half the time?
It is clear that both magnetic flux and angular momentum must be
redistributed during collapse to produce stars with reasonable
fields and rotation rates, and these processes will affect
the formation of binaries and protoplanetary disks.
Some of these questions will be addressed in the next sections on
globules and cores, while others will be discussed in the context of
testing specific theories.

\subsection{Globules and Cores: Overall Properties}         \label{globules}

Nearby small dark clouds, or globules, are natural places to look for
isolated star formation (Bok \& Reilly 1947). A catalog of 248 globules
(Clemens \& Barvainis 1988) has provided the basis for many studies.
Yun \& Clemens (1990) found that 23\% of the CB globules
appear to contain embedded infrared sources, with \sed s typical of
star forming regions (Yun 1993). About one-third of the globules
with embedded sources have evidence of outflows (Yun \& Clemens 1992;
Henning \& Launhardt 1998). Clearly, star formation does occur in isolated
globules.

Within the larger dark clouds, one can identify numerous regions of high
opacity (e.g. Myers et al 1983), commonly called cores (Myers 1985).
Surveys of such regions in low-excitation lines of \ammonia\
(e.g. Benson \& Myers 1989) led to
the picture of an isolated core within a larger cloud,
which then might pursue its course
toward star formation in relative isolation from the rest of the cloud.
Most intriguing was the fact that the \ammonia\ linewidths in many of these
cores indicated that the turbulence was subsonic (Myers 1983);
in some cores, thermal broadening  of \ammonia\ lines even dominated
over turbulent broadening (Myers \& Benson 1983, Fuller \& Myers 1993).
Although later studies in other lines indicated a more
complex dynamical situation (Zhou et al 1989, Butner et al 1995),
the \ammonia\ data provided observational support for theories describing
the collapse of isothermal spheres (Shu 1977).
The discovery of \iras\ sources in half
of these cores (Beichman et al 1986) indicated that they were indeed
sites of star formation.
The observational and theoretical developments were synthesized into an
influential paradigm
for low mass star formation (Shu et al 1987a, \S \ref{theorydet}).

Globules would appear to be an ideal sample for measuring sizes since
the effects of the environment are minimized by their isolation, but
distances are uncertain. Based on the angular size of the optical images
and an assumed average distance of 600 pc (Clemens \& Barvainis 1988),
the mean size, $\meanl = 0.7$ pc.
A subsample of these with distance estimates were mapped in molecular lines,
yielding much smaller average sizes: $\meanl = 0.33\pm 0.15$ pc for
a sample of 6 ``typical" globules mapped in CS (Launhardt et al 1998);
maps of the same globules in C$^{18}$O \jj21 (Wang et al 1995) give sizes
smaller by a factor of $2.9\pm1.6$.
A sample of 11 globules in the southern sky, mapped in \ammonia\
(Bourke et al 1995) have $\meanl = 0.21 \pm 0.08$ pc.

Cores in nearby dark clouds have the advantage of having well-determined
distances; the main issue is how clearly they stand out from the bulk
of the molecular cloud.  Gregersen (1998) found that some known cores
are barely visible above the general cloud emission in the \cooo\ \jj10\ line.
The mean size of a sample of 16 cores mapped in \ammonia\ is 0.15 pc,
whereas CS \jj21\ gives 0.27 pc, and C$^{18}$O \jj10\ gives
0.36 pc (Myers et al 1991). These differences may
reflect the effects of opacity, chemistry, and density structure.

Globules are generally not spherical. By fitting the opaque cores with
ellipses, Clemens \& Barvainis (1988) found a mean aspect ratio
($a/b$) of 2.0. Measurements of aspect ratio in
tracers of reasonably dense gas toward globules
give $\meanar \sim 1.5-2$ (Wang et al 1995, Bourke et al 1995), and
cores in larger clouds have $\meanar \sim 2$ (Myers et al 1991).
Myers et al and Ryden (1996) have argued that the underlying
3-D shapes were more likely to be prolate, with axial ratios around 2,
than oblate, where axial ratios of 3--10 were needed. However,
toroids may also be able to match the data because of their
central density minima (Li \& Shu 1996).

The uncertainties in size are reflected directly into uncertainties
in mass.
For the sample of globules mapped by Bourke et al (1995) in \ammonia,
$\mean\mc = 4\pm1$ \msun, compared to $10\pm2$ \msun\ for cores. Larger
masses are obtained from other tracers. For the sample of globules
studied by Launhardt et al (1998), $\mean\mv = 26\pm12$ based on CS \jj21
and $10\pm6$, based on \cooo\ \jj21.
Studies of the cores in larger clouds found similar ranges and differences
among tracers (e.g. Fuller 1989, Zhou et al 1994a).
A series of studies of the Taurus cloud complex
has provided an unbiased survey of cores with known distance identified by
\cooo\ \jj10\ maps.  Starting from a large-scale map of \coo\ \jj10\
(Mizuno et al 1995), Onishi et al (1996) covered 90\% of the area with
$N > 3.5\ee{21}$ with a map of \cooo\ \jj10\ with 0.1 pc resolution.
They identified 40 cores with $\meanl = 0.46$ pc, $\meanar = 1.8$, and
$\mean\mc = 23$ \msun. The sizes extended over a range of a factor of 6
and masses over a factor of 80. Comparing these cores to the distribution
of T Tauri stars, infrared sources, and \hcopi\ emission, Onishi et al
(1998) found that all cores with $N > 8\ee{21}$ \cmc\ are associated with
\hcopi\ emission and/or cold \iras\ sources. In addition,
the larger cores always contained multiple objects. They
concluded that the core mass per star-forming event is relatively
constant at 11 \msun.

It is clear that characterizing globules and cores by a typical size
and mass is an oversimplification. First, they come in a range of sizes
that is probably just the low end of the general distribution of cloud sizes.
Second, the size and mass depend strongly on the tracer and method used
to measure them. If we ignore these caveats, it is probably fair to say
that most of these regions have sizes measured in tracers of reasonably
dense gas in the range of a few tenths of a pc and masses of less than
100 \msun, with more small, low mass cores than massive ones. The larger cores
tend to be fragmented, so that the mass of gas with $n \geq \eten4$ \cmv\
tends toward 10 \msun, within a factor of 2. Star formation may occur when
the column density exceeds 8\ee{21} \cmc, corresponding in this case to
$\n \sim \eten4$ \cmv.

\subsection{Globules and Cores: Internal Conditions}      \label{internal}


If the sizes and masses of globules and cores are poorly defined,
at least the temperatures seem well understood.
Early CO observations (Dickman 1975)
showed that the darker globules are cold (\tk $ \sim 10$ K),
as expected for regions with only cosmic ray heating.  Similar results were
found from \ammonia\ (Martin \& Barrett 1978 Myers \& Benson 1983;
Bourke et al 1995).  Clemens et al (1991) showed the distribution of CO
temperatures for a large sample of globules; the main peak corresponded
to $\tk = 8.5$ K, with a small tail to higher \tk.
Determination of the dust temperature was more difficult,
but Keene (1981) measured $\td = 13-16$ K in B133, a starless core.
More recently, Ward-Thompson et al (1998) used
\iso\ data to measure $\td = 13$ K in another starless core.
These results are similar to predictions for cores heated by the
interstellar radiation field  (Leung 1975), though
\td\ is expected to be lower in the deep interiors.

A sample of starless globules with $\av \sim 1-2$ mag
produced no detections of \ammonia\ (Kane et al 1994), but
surveys of \form\ (Wang et al 1995) and CS (Launhardt et al 1998, Henning
\& Launhardt 1998)
toward more opaque globules indicate that dense gas is present in some.
The detection rate of CS \jj21\ emission was much higher
in globules with infrared sources.
Butner et al (1995) analyzed multiple transitions of \dcop\ in 18 low-mass
cores, finding $\mean{\log n (\cmv)} \simeq 5$, with a tendency to slightly
higher values in the cores with infrared sources. Thus, gas still denser
than the $\mean{\log n (\cmv)} \simeq 4$ gas traced by \ammonia\ exists in
these cores.

Discussion of the kinematics of globules and cores has often focused on
the relationship between linewidth (\dv) and size ($l$ or $R$). This
relation is much less clearly established for cores within a single
cloud than is the relation for clouds as a whole (\S \ref {probekin}),
and it may have a different origin (Myers 1985).
Goodman et al (1998) have distinguished four types of
linewidth-size relationships. Most studies have employed Goodman Type 1
relations (multi-tracer, multi-core), but it is difficult to distinguish
different causes in such relations. Goodman Type 2 relations (single-tracer,
multi-core) relations within a single cloud would reveal whether the virial
masses of cores are reliable. Interestingly, the most systematic
study of cores in a single cloud found no correlation in 24 cores in
Taurus mapped in \cooo\ \jj10\ ($\gamma = 0.0\pm0.2$), but the range of sizes
(0.13 to 0.4 pc) may have been insufficient (Onishi et al 1996).

To study the kinematics of individual cores, the most useful relations are
Goodman Types 3 (multi-tracer, single-core) and 4 (single-tracer, single-core).
In these, a central position is defined, either by
an infrared source or a line peak. A Type 3 relationship using \ammonia, \cooo\
\jj10\ and CS \jj21\ lines was explored by Fuller \& Myers (1992), who found
$\dv \propto R^{\gamma}$, with $R$ the radius of the half-power contour.
Caselli \& Myers (1995) added \coo\ \jj10\ data and constructed a Type 1
relation for 8 starless cores, after removing the thermal broadening.
The mean $\gamma$ was $0.53\pm0.07$ with a correlation coefficient of 0.81.
However, both these
relationships depend strongly on the fact that the \ammonia\ lines
have small \dv\ and $R$. Some other species (e.g. HC$_3$N,
Fuller \& Myers 1993) also have narrow lines and small sizes, but
\dcop\ emission has much wider lines over a similar map size (Butner et al
1995), raising the possibility that chemical effects cause different molecules
to trace different kinematic regimes within the same overall core.
Goodman et al (1998) suggest
that the \dcop\ is excited in a region outside the \ammonia\ region and
that the size of the \dcop\ region is underestimated. On the other hand,
some chemical simulations indicate that \ammonia\ will deplete in dense
cores while ions like \dcop\ will not (Rawlings et al 1992),
suggesting the opposite solution.

A way to avoid such effects is to use a Type 4
relation, searching for a correlation between $\dv_{NT}$ as spectra are
averaged in rings of larger size, though line of sight confusion cannot be
avoided, as one gives up the ability to tune the density sensitivity.
This method has almost never been applied to tracers of dense gas,
but Goodman et al (1998) use an indirect method to obtain a Type 4 relation
for three clouds in OH, \cooo, and \ammonia. The relations are very flat
($\gamma = 0.1$ to 0.3), and $\gamma \neq 0$ has statistical significance only
for OH, which traces the least dense gas. In particular, \ammonia\
shows no significant Type 4 relation, having narrow lines on every scale
(see also Barranco \& Goodman 1998). Goodman et al (1998) interpret these
results in terms of a ``transition to coherence" at the scale of 0.1--0.2
pc from the center of a dense core. Inside that radius, the turbulence
becomes subsonic and no longer decreases with size (Barranco \& Goodman 1998).
While this picture accords nicely with the idea that cores can be
distinguished from the surroundings and treated as ``units" in low-mass
star formation, the discrepancy between values of \dv\ measured in different
tracers of the dense core (cf Butner et al 1995) indicate that caution
is required in interpreting the \ammonia\ data.

Rotation can be detected in some low-mass cores, but the ratio of rotational
to gravitational energy has a typical value of 0.02 on scales of 0.1 pc
(Goodman et al 1993). The inferred rotation axes are not correlated with
the orientation of cloud elongation, again suggesting that rotation is not
dynamically important on the scale of 0.1 pc. Ohashi et al (1997a) find
evidence in several star-forming cores for a transition at $r \sim 0.03$ pc,
inside of which the specific
angular momentum appears to be constant at about $\eten{-3}$ \kms pc down
to scales of about 200 AU.

Knowledge of the magnetic field strength in cores and globules would be
extremely valuable in assessing whether they are subcritical or supercritical.
Unfortunately, Zeeman measurements of these regions are extremely difficult
because they must be done with emission, unless there is a chance
alignment with a background radio source. Crutcher et al (1993) detected
Zeeman splitting in OH in only one of 12 positions in nearby clouds.
Statistical analysis of the detection and the upper limits, including
the effects of random orientation of the field, led to the conclusion that
the data could not falsify the hypothesis that the clouds were subcritical.
Another problem is that the OH emission probes relatively low densities
($n \sim \eten3$ \cmv). Attempts to use CN to probe denser gas produced
upper limits that were less than expected for subcritical clouds, but the
small sample size and other uncertainties again prevented a definitive
conclusion (Crutcher et al 1996). Improved sensitivity and larger samples
are crucial to progress in this field. At present,
no clear examples of subcritical cores have been found (Crutcher 1999a),
but uncertainties are sufficient to allow this possibility.

Onishi et al (1996) found that the major axis of the cores they identified
in Taurus tended to be perpendicular to the optical polarization vectors and
hence \bperp. Counterexamples are known in other regions (e.g. Vrba et al 1976)
and the fact that optical polarization does not trace the dense portions
of clouds (Goodman 1996) suggests that this result be treated cautiously.
Further studies of \bperp\ using dust emission at long wavelengths
are clearly needed.

The median ionization fraction of 23 low-mass cores is 9\ee{-8},
with a range of log$x_e$ of $-7.5$ to $-6.5$, with typical
uncertainties of a factor of 0.5 in the log (Williams et al 1998).
Cores with stars do not differ significantly in $x_e$ from cores
without stars, consistent with cosmic ray ionization. For a cloud with
$ n = \eten4$ \cmv, the ambipolar diffusion timescale, $t_{AD} \sim
7 \ee6 {\rm yr} \sim 20 t_{ff}$. If the cores are subcritical, they
will evolve much more slowly than a free fall time. Recent comparisons
of the line profiles of ionized and neutral species have been interpreted
as setting an upper limit on the ion-neutral drift velocity of $0.03$
\kms, consistent with that expected from ambipolar diffusion (Benson
et al 1998).

To summarize the last two subsections,
there is considerable evidence that distinct cores can
be identified, both as isolated globules and as cores within larger
clouds. While there is a substantial range of properties, scales of
0.1 pc in size and 10 \msun\ in mass seem common. The cores are cold
($\tk \sim \td \sim 10$ K) and contain gas with $\n \sim \eten4$ \cmv,
extending in some cases up to $\n \sim \eten5$ \cmv. While different molecules
disagree about the magnitude of the effect, these cores seem to be regions
of decreased turbulence compared to the surroundings. While no clear cases
of subcritical cores have been found, the hypothesis that low-mass
stars form in subcritical cores cannot be ruled out observationally.
How these cores form is beyond the scope of this review, but again
one can imagine two distinct scenarios: ambipolar diffusion brings an
initially subcritical core to a supercritical state; or dissipation of
turbulence plays a similar role in a core originally supported by
turbulence (Myers \& Lazarian 1998). In the latter
picture, cores may build up from accretion of smaller diffuse elements
(Kuiper et al 1996), perhaps the structures inferred by Falgarone et al (1998).

\subsection{Classification of Sources and Evolutionary Scenarios}
\label{classes}

The \iras\ survey provided spectral energy distributions over
a wide wavelength range for many cores (e.g. Beichman et al 1986),
leading to a classification scheme for infrared sources.
In the original scheme (Lada \& Wilking 1984, Lada 1987),
the spectral index between 2 \um\ and the longest observed wavelength
was used to divide sources into three Classes, designated by
roman numerals, with Class I indicating the most emission at long wavelengths.
These classes rapidly became identified with stages in the
emerging theoretical paradigm (Shu et al 1987a):
Class I sources are believed to be undergoing infall with simultaneous
bipolar outflow, Class II sources are typically visible T Tauri stars with
disks and winds, and Class III sources have accreted or dissipated most of
the material, leaving a pre-main-sequence star, possibly with planets
(Adams et al 1987; Lada 1991).

More recently, \submm\ continuum observations have revealed
a large number of sources with emission peaking at still longer wavelengths.
Some of these new sources also have infrared sources and powerful bipolar
outflows, indicating that a central object has formed; these have been
designated Class 0 (Andr\'e et al 1993). Andr\'e \& Montmerle (1994) argued
that Class 0 sources represent the primary infall stage, in which there is
still more circumstellar than stellar matter.
Outflows appear to be most intense in the
earliest stages, declining later (Bontemps et al 1996).
Other cores with \submm\ emission have no \iras\ sources
and probably precede the formation of a central object. These were found
among the ``starless cores" of Benson \& Myers (1989),
and Ward-Thompson et al (1994)
referred to them as ``pre-protostellar cores." The predestination implicit
in this name has made it controversial, and I will use the less descriptive
(and somewhat tongue-in-cheek) term, Class $-1$. There has also been some
controversy over whether Class 0 sources are really distinct from Class I
sources or just more extreme versions. The case for Class 0 sources as
a distinct stage can be found in Andr\'e et al (2000).

While classification has an honored history in astronomy, serious tests of
theory are facilitated by continuous variables. Myers \& Ladd (1993) suggested
that we characterize the \sed\ by the flux-weighted mean frequency, or more
suggestively, by the temperature of a black body with the same mean frequency.
The latter (\tbol) was calculated by Chen et al (1995) for many sources and
the following boundary lines in \tbol\ were found to coincide with the
traditional classes: $\tbol < 70$K for Class 0, $70 \leq \tbol \leq 650$K for
Class I, and $650 < \tbol \leq 2800$K for Class II.
In a crude sense, \tbol\ captures the ``coolness"
of the \sed, which is related to how opaque the dust is,
but it can be affected strongly by how much \mir\ and \nir\ emission
escapes, and thus by geometry. Other measures, such as the ratio of emission
at a \submm\ wavelength to the bolometric luminosity ($L_{smm}/L_{bol}$), may
also be useful. One of the problems with all these measures is that the bulk
of the energy for classes earlier than II emerges at \fir\ wavelengths, where
resolution has been poor. Maps of \submm\ emission are showing that  many
\nir\ sources are displaced from the \submm\ peaks and may have been falsely
identified as Class I sources (Motte et al 1998).
Ultimately, higher spatial resolution in the \fir\
will be needed to sort out this confusion.

\subsection{Detailed Theories}                      \label{theorydet}

The recent focus on the formation of isolated, low-mass stars is at least
partly due to the fact that it is more tractable theoretically than the
formation of massive stars in clusters. Shu (1977) argued that collapse begins
in a centrally condensed configuration and propagates outward at the sound
speed ($a$); matter inside $\rinf = at$ is infalling after a time $t$.
Because the collapse in this model is self-similar, the structure
can be specified at any time from a single solution.
This situation is called inside-out collapse. In Shu's picture, the
pre-collapse
configuration is an isothermal sphere, with $n(r) \propto r^{-p}$ and $p=2$.
Calculations of core formation via ambipolar diffusion gradually approach a
configuration with an envelope that is close to a power law, but with a core
of ever-shrinking size and mass	where $p\sim 0$ (e.g. Mouschovias 1991).
It is natural to identify the ambipolar diffusion stage with the
Class $-1$ stage, and the inside-out collapse with Class 0 to Class I.
As collapse proceeds, the material inside \rinf\
becomes less dense, with a power law approaching $p=1.5$ in the inner regions,
after a transition region where the density is not a power law.
In addition, $v(r) \propto r^{-0.5}$ at small $r$.

There are many other solutions to the collapse problem (Hunter 1977,
Foster and Chevalier 1993) ranging from the
inside-out collapse to overall collapse (Larson 1969, Penston 1969).
Henriksen et al (1997) have argued that collapse begins before the
isothermal sphere is established; the inner ($p=0$) core undergoes a
rapid collapse that they identify with Class 0 sources. They suggest that
Class I sources represent the inside-out collapse phase, which appears only
when the wave of infall has reached the $p=2$ envelope.

Either rotation or magnetic fields will break spherical symmetry.
Terebey et al (1984) added slow rotation of the original cloud at
an angular velocity of $\Omega$ to the
inside-out collapse picture, resulting in another characteristic radius,
the point where the rotation speed equals the infall speed.
A rotationally supported disk should form somewhere inside this centrifugal
radius (Shu et al 1987a)

\begin{equation}
r_c = {{G^3 M^3 \Omega^2} \over {16 a^8}},
\end{equation}
where $M$ is the mass already in the star and disk. Since
disks are implicated in all models of the ultimate source of the outflows,
the formation of the disk may also signal the start of the outflow. Once
material close to the rotation axis has accreted (or been blown out) all
further
accretion onto the star should occur through the disk.

Core formation in a magnetic field should produce a flattened structure
(e.g. Fiedler \&  Mouschovias 1993) and Li and Shu (1996) argue that the
equilibrium
structure equivalent to the isothermal sphere is the isothermal toroid.
Useful insights into the collapse of a magnetized cloud have resulted
from calculations in spherical geometry (Safier et al 1997, Li 1998) or
for thin disks (Ciolek \& K\"onigl 1998). Some calculations of collapse in
two dimensions have been done (e.g. Fiedler \& Mouschovias 1993).
A magnetically-channeled, flattened structure may appear; this has been called
a pseudo-disk (Galli \& Shu 1993a, 1993b) to distinguish it from the
rotationally supported disk. In this picture, material would flow into a
pseudo-disk at a scale of $\sim 1000$ AU before becoming rotationally supported
on the scale of $r_c$.
The breaking of spherical symmetry on scales of 1000 AU
may explain some of the larger structures seen in some regions (see
Mundy et al 2000 for a review).

Ultimately, theory should be able to predict the conditions that lead to
binary formation, but this is not presently possible. Steps toward this
goal can be seen in numerical calculations of collapse with rotation
(e.g. Bonnell \& Bastien 1993, Truelove et al 1998, Boss 1998).
Rotation and magnetic fields have been combined in a series of calculations
by Basu \& Mouschovias (1995 and references therein); Basu (1998) has
considered the effects of magnetic fields on the formation of rotating disks.

Theoretical models make predictions that are testable, at least in principle.
In the simplest picture of the inside-out collapse of the rotationless,
non-magnetic isothermal sphere, the theory predicts all the density and
velocity structure with only the sound speed as a free parameter.
Rotation adds $\Omega$ and magnetic fields add a reference field or
equivalent as additional parameters. Departure from spherical symmetry
adds the additional observational parameter of viewing angle. Consequently,
observational tests have focused primarily on testing the simplest models.

\subsection{Tests of Evolutionary Hypotheses and Theory} \label{tests}

Both the empirical evolutionary sequence based on the class system
and the detailed theoretical predictions can be tested by detailed
observations.
One can compute the expected changes in the continuum emission as a function
of time for a particular model.
Examples include  plots of $L$ versus \av\ (Adams 1990), $L$ versus
\tbol\ (Myers et al 1998), and $L_{mm}$ versus $L$ (Saraceno et al 1996).
For example, as time goes on, \av\ decreases, \tbol\ increases, and $L$
reaches a peak and declines.
At present the models are somewhat idealized and simplified, and different
dust opacities need to be considered.
Comparison of the number of objects observed in various parts of these
diagrams with those expected from lifetime considerations can provide an
overall check on evolutionary scenarios and provide age estimates for
objects in different classes (e.g. Andr\'e et al 2000).
One can also test the models against observations of particular objects.
However, models of source structure constrained only by the \sed\ are not
unique (Butner et al 1991, Men'shchikov \& Henning 1997).
Observational determination of $n(r)$ and $v(r)$ can apply more stringent
tests.

Maps of continuum emission from dust can trace the column density as a
function of radius quite effectively, if the temperature distribution is known.
With an assumption about geometry, this information can be related to $n(r)$.
For the Class $-1$ sources, with no central
object, the core should be isothermal ($\td \sim 5-10$ K) or warmer on the
outside if exposed to the interstellar radiation field (Leung 1975, Spencer
\& Leung 1978). New results from \iso\ will put tighter limits on possible
internal energy sources (e.g. Ward-Thompson et al 1998, Andr\'e et al 2000).
Based on small maps of \submm\ emission, Ward-Thompson et al (1994) found
that Class $-1$ sources were not characterized
by single power laws in column density; they fit the distributions with
broken power laws, indicating a shallower distribution closer to the center.
This distribution appears consistent with the interpretation that these
are cores still forming by ambipolar diffusion.
Maps at 1.3 mm (Andr\'e et al 1996, Ward-Thompson et al
1999) confirm the early results: the column density is quite constant in an
inner region ($r < 3000$ to 4000 AU), with $M \sim 0.7$ \msun.
SCUBA maps of these sources are just becoming available, but they suggest
similar conclusions (D Ward-Thompson, personal communication,  Shirley et al
1998).
In addition, some of the cores are filamentary and fragmented (Ward-Thompson
et al 1999). Citing these observed properties, together with a statistical
argument regarding lifetimes, Ward-Thompson et al (1999) now argue that
ambipolar diffusion in subcritical
cores does not match the data (see also Andr\'e et al 2000).

For cores with central sources (Class $> -1$), $\td(r)$ will decline with
radius from the source. If the emission is in the Rayleigh-Jeans limit,
$I_{\nu} \propto \td \opacity N$. If \opacity\ is not a function of $r$,
$I_{\nu}(\theta) \propto \int \td (r) n(r) dl$, where the integration
is performed along the line of sight ($dl$).
If one avoids the central beam and any outer cut-off,
assumes the optically thin expression for
$\td (r) \propto r^{-q}$, and fits a power law to $I_{\nu}(\theta) \propto
\theta^{-m}$, then $p = m + 1 - q$ (Adams 1991, Ladd et al 1991).
With a proper calculation of $\td(r)$
and convolution with the beam, one can include all the information (Adams
1991).
This technique has been applied in the \fir\ (Butner et al 1991), but it
is most useful at longer wavelengths. Ladd et al (1991) found $p = 1.7\pm 0.3$
in two cores, assuming $q = 0.4$, as expected for $\beta = 1$ (equation 14).
Results so far are roughly consistent with theoretical models, but
results in this area from the new \submm\ cameras
will explode about the time this review goes to press.

Another issue arises for cores with central objects. If a circumstellar
disk contributes
significantly to the emission in the central beam, it will increase the
fitted value of $m$. Disks contribute more importantly at longer wavelengths,
so the disk contribution is more important at \mm\ wavelengths (e.g. Chandler
et al 1995). Luckily, interferometers are available
at those wavelengths, and observations with a wide range of antenna spacings
can separate the contributions of a disk and envelope.
Application of this technique by Looney et al (1997) to L1551~IRS5,
a Class I source, reveals very complex structure (Figure 3):
binary circumstellar disks (cf Rodr\'{\i}guez et al 1998),
a circumbinary structure (perhaps a pseudo-disk), and an envelope with
a density distribution consistent with a power law ($p = 1.5$ to 2).
This technique promises to be very fruitful in tracing the
flow of matter from envelope to disk. Early results indicated that disks are
more prominent in more evolved (Class II) systems (Ohashi et al 1996), but
compact structures are detectable in some younger systems (Terebey et al 1993).
Higher resolution observations and careful analysis will be needed to
distinguish envelopes, pseudo-disks, and Keplerian disks (see Mundy et al
2000 for a review). At the moment, one can only say that disks in the
Class 0 stage are not significantly {\it more} massive than disks in
later classes (Mundy et al 2000). Meanwhile, the interferometric data
confirm the tendency of envelope mass to decrease with class number inferred
from single-dish data (Mundy et al 2000).

\setcounter{figure}{2}
\begin{figure}[ht!]
\centering
\vspace*{11.5cm}
   \leavevmode
   \includegraphics{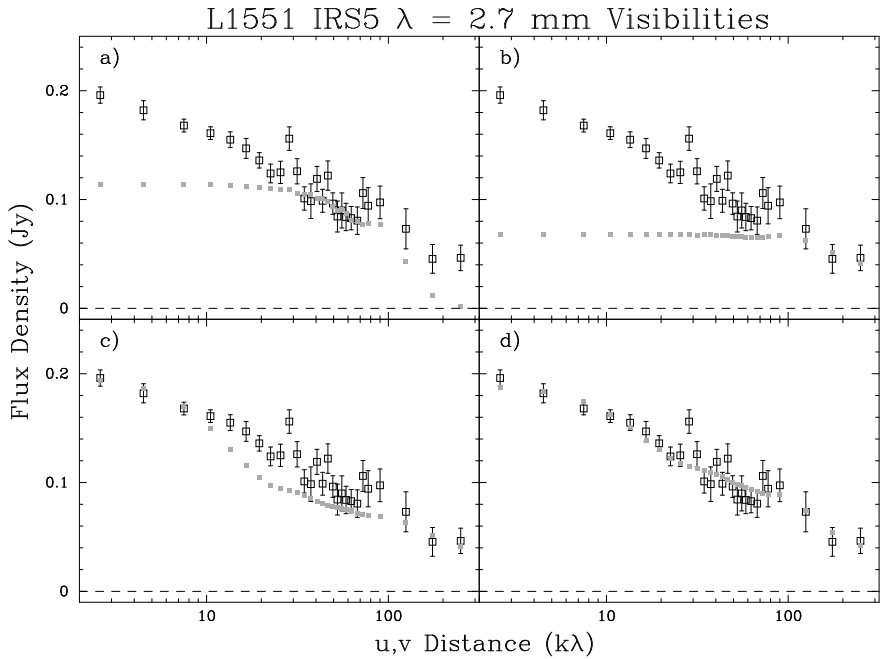}
\caption{ Observed visibilities of L1551 IRS 5 at 2.7 mm, binned in annuli
(open
squares with error bars), are plotted versus the projected baseline in units
of $\eten3$ times the wavelength. Different models are shown in each panel by
the
small boxes. Panel a has a model with only a Gaussian source of radius 80 AU;
panel b has a model with two point sources constrained to match the map; panel
c adds a truncated power law ($p = 1.5$, $q = 0.5$) envelope to the two point
sources; panel d adds to the previous components a circumbinary structure,
represented by a Gaussian. In the final optimization (panel d),
the envelope has a mass of 0.28 \msun\ and an outer radius of 1100 AU, the
circumbinary structure has a mass of 0.04 \msun, and the circumstellar
disk masses are 0.024 and 0.009 \msun\ (Looney et al 1997).
}
\end{figure}
Similar techniques have been used for maps of molecular line emission.
For example, \coo\ emission has been used to trace column density in
the outer regions of dark clouds. With an assumption of spherical symmetry,
the results favor $p \sim 2$ in most clouds (Snell 1981, Arquilla \& Goldsmith
1985). The \coo\ lines become optically thick in the inner regions;
studies with higher spatial resolution in rarer isotopomers, like \cooo\
or \coooo, tend to show somewhat more shallow density distributions
than expected by the standard model (Zhou et al 1994b, Wang et al 1995).
Depletion in the dense, cold cores may still confuse matters (e.g.
Kuiper et al 1996, \S \ref{probecol}).
Addressing the question of evolution, Ladd et al (1998) used two
transitions of \cooo\ and \coooo\ to show that $N$ toward the central
source declines with \tbol, with a power between $0.4$ and $1.0$.
To reproduce the inferred rapid decrease in mass with time, they
suggest higher early mass loss than predicted by the standard model.

By observing
a series of lines of different critical density, modeling those lines with
a particular cloud model and appropriate radiative transport, and predicting
the emission into the beams used for the observations, one can constrain the
run of density more directly.  Studies using two transitions of \form\
have again supported $p = 2 \pm 0.5$ on relatively large scales (Loren et al
1983, Fulkerson \& Clark 1984). When interferometery of \form\ was used to
improve the resolution on one core, $p$ appeared to decrease at small $r$
(Zhou et al 1990), in agreement with the model of Shu (1977).
Much of the recent work on this topic has involved testing of detailed collapse
models, including velocity fields and the complete density law, rather than
a single power law, as described in the next section.

\subsection{Collapse}                                \label{collapse}

The calculation of line profiles as a function of time (Zhou 1992) for the
collapse models of Shu (1977) and Larson (1969) and Penston (1969),
along with claims of collapse in a low-mass star forming region
(Walker et al 1986), reinvigorated the study of protostellar collapse.
Collapsing clouds
will depart from the linewidth-size relation (\S \ref{internal}),
having systematically larger linewidths for a given size (Zhou 1992).
Other simulations of line profiles range from a simple two-layer model
(Myers et al 1996) to detailed calculations of radiative transport
(Choi et al 1995, Walker et al 1994, Wiesemeyer 1997, 1999).

Zhou et al (1993) showed that several lines of CS and \form\ observed towards
B335, a globule with a Class 0 source, could be fitted very well using the
exact $n(r)$ and
$v(r)$ of the inside-out collapse model. Using a more self-consistent radiative
transport code, Choi et al (1995) found slightly different best-fit parameters.
Using a sound speed determined from lines away from the collapse region,
the only free parameters were the time since collapse began and the abundance
of
each molecule. With several lines of each molecule, the problem is quite
constrained (Figure 4). This work was important in
gaining acceptance for the idea that collapse had finally been seen.

\begin{figure}[ht!]
\centering
\vspace*{11.5cm}
   \leavevmode
   \includegraphics{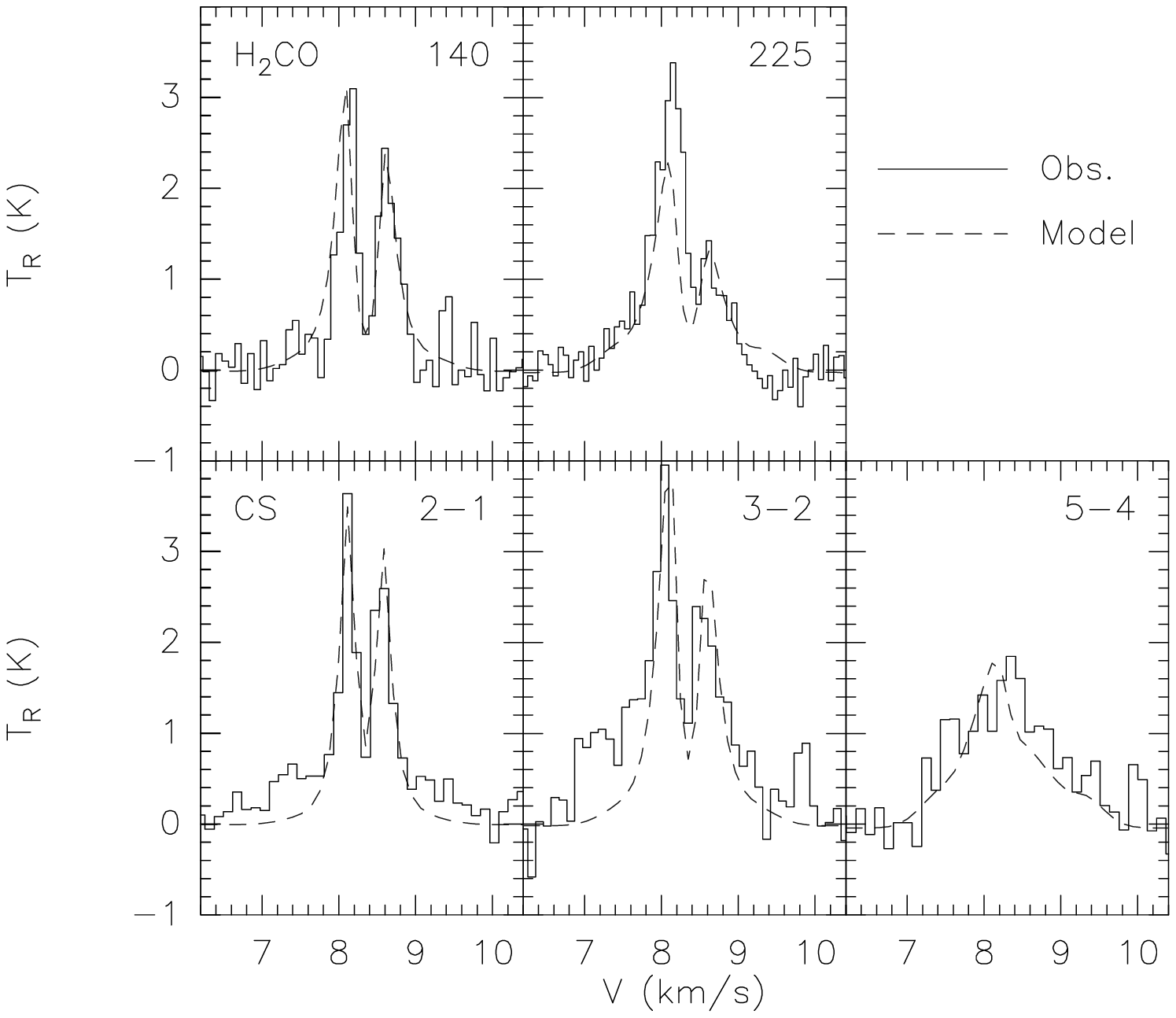}

\caption{ Line profiles of \form\ and CS emission (solid histograms)
toward B335 and the best-fitting model (dashed lines). The model line
profiles were calculated with a Monte Carlo code, including non-LTE
excitation and
trapping, with an input density and velocity field taken directly from
the collapse model of Shu (1977) and a temperature field calculated with
a separate dust radiation transport code. The best-fitting model has
an infall radius of 0.03 pc (Choi et al 1995).
}
\end{figure}

Examination of the line profiles in Figure 4 reveals that most are strongly
self-absorbed. Recall that the overall collapse idea of Goldreich and Kwan
(1974) was designed to avoid self-absorbed profiles. The difference is that
Goldreich and Kwan assumed that $v(r) \propto r$, so that every velocity
corresponded
to a single point along the line of sight. In contrast, the inside-out collapse
model predicts $v(r) \propto r^{-0.5}$ inside a static envelope. If the line
has substantial opacity in the static envelope, it will produce a narrow
self-absorption at the velocity centroid of the core (Figure 5).
The other striking feature of the spectra in Figure 4 is
that the blue-shifted peak is stronger than the red-shifted peak. This ``blue"
profile occurs because the $v(r) \propto r^{-0.5}$ velocity field has
two points along any line of sight with the same Doppler shift (Figure 6).
For a centrally peaked temperature and density distribution, lines
with high critical densities will have higher \tex\ at the point closer to the
center. If the line has sufficient opacity at the relevant
point in the cloud, the high \tex\ point in the red peak will be obscured by
the
lower \tex\ one, making the red peak weaker than the blue peak (Figure 6).
Thus a collapsing
cloud with a velocity and density gradient similar to those in the inside-out
collapse model will produce blue profiles in lines with suitable excitation and
opacity properties. A double-peaked profile with a stronger blue peak or
a blue-skewed profile relative to an optically thin line then becomes a
signature of collapse. These features were discussed by Zhou \& Evans (1994)
and, in a more limited context, by Snell \& Loren (1977) and
Leung \& Brown (1977).

\begin{figure}[ht!]
\centering
\vspace*{11.5cm}
   \leavevmode
   \includegraphics{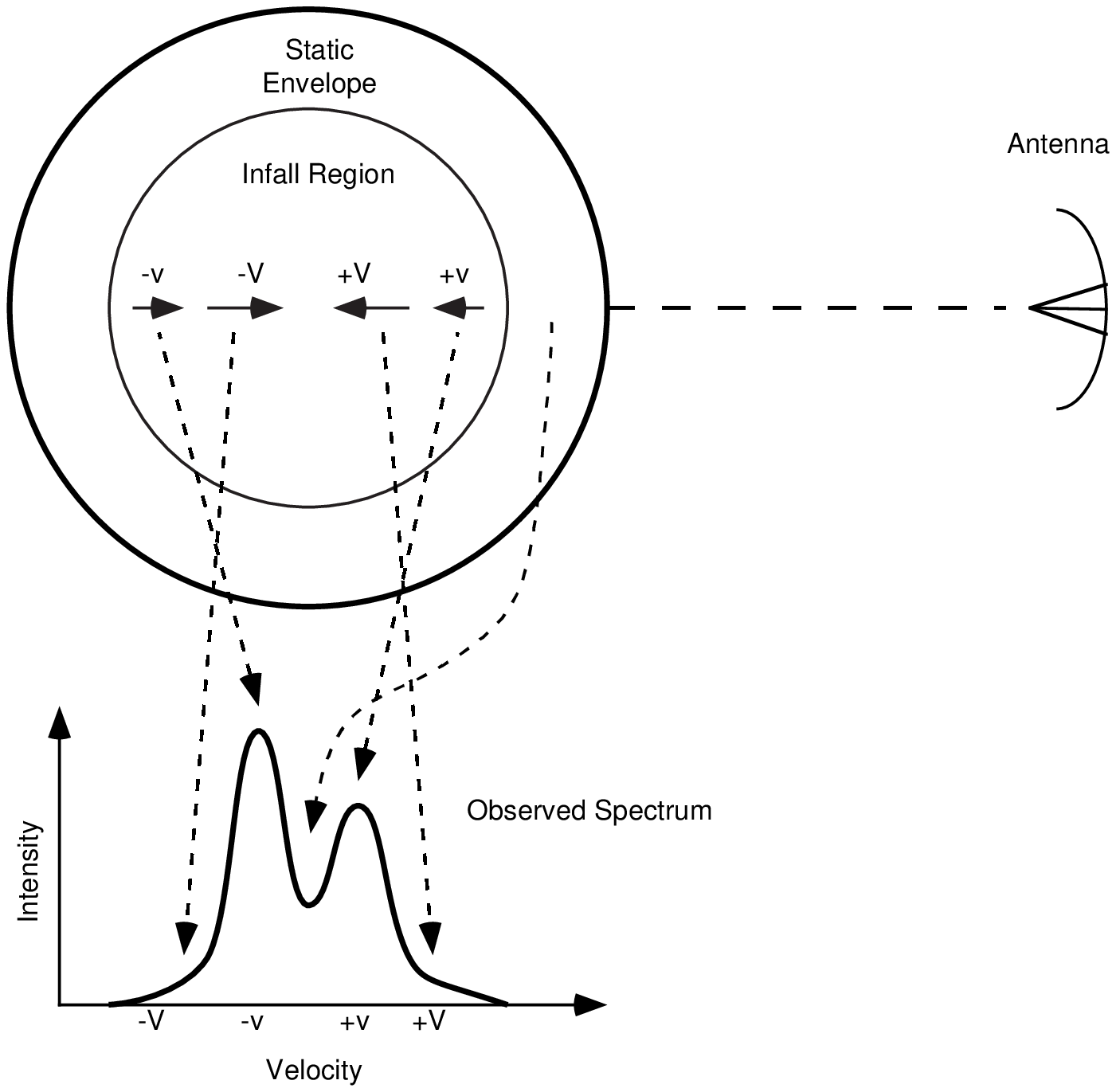}

\caption{ The origin of various parts of the line profile for a cloud
undergoing inside-out collapse. The static envelope outside \rinf\
produces the central self-absorption dip, the blue peak comes from the
back of the cloud, and the red peak from the front of the cloud.
The faster collapse near the center produces line wings, but these are
usually confused by outflow wings.
}

\end{figure}

Of course, the collapse interpretation of a blue profile is not unique.
Such profiles can be produced in a variety of ways.
To be a plausible candidate for collapse, a core must also show these features:
an optically thin line must peak between the two peaks of the opaque line;
the strength and skewness should peak on the central source;
and the two peaks should not be caused by clumps in an outflow.
The optically thin line is particularly crucial,
since two cloud components, for example, colliding fragments,
could produce the double-peaked blue profile, but they would also produce a
double-peaked profile in the optically thin line.

Rotation, combined with self-absorption, can create a line profile like
that of collapse (Menten et al 1987, Adelson \& Leung 1988), but toward
the center of rotation,
the line would be symmetric (Zhou 1995). Rotating collapse can cause the
line profiles to shift from blue to red-skewed on either side of the rotation
axis, with the sign of the effect depending on how the rotation varies with
radius (Zhou 1995). Maps of the line centroid can be used to separate rotation
from collapse (Adelson \& Leung 1988, Walker et al 1994).

To turn a collapse candidate into
a believable case of collapse, one has to map the line profiles, account for
the effects of outflows, model rotation if present, and show that a collapse
model fits the line profiles. To date this has been done only for a few
sources:
B335 (Zhou et al 1993, Choi et al 1995), L1527 (Myers et al 1995, Zhou et al
1996, Gregersen et al 1997), and \iras\/ 16293 (Zhou 1995,
Narayanan et al 1998).  Of this group,
only \iras\/ 16293, rotating about 20 times faster than B335, is known
to be a binary (Wootten 1989, Mundy et al 1992), supporting the idea
that faster rotation is more likely to produce a binary.
Mathieu (1994) reviews binarity in the pre-main-sequence stage, and Mundy
et al (2000) discuss recent evidence on the earlier stages.

Interferometric observations have also revealed infall motions and rotational
motions on scales of $\sim 1000$ AU in several sources (e.g. Ohashi et
al 1997b; Momose et al 1998). Such studies can reveal how matter makes the
transition from infall to a rotating disk.
Inevitably, irregularities in the density and velocity fields will confuse
matters in real sources, and these may be more noticeable with interferometers.
Outflows are particularly troublesome (Hogerheijde et al 1998).
Extreme blue/red ratios are seen in interferometric observations of
\hcop\ and HCN \jj10\ lines, which are difficult to reproduce with
standard models (Choi et al 1999).
Even in B335, the best case for collapse, Velusamy et al (1995)
found evidence for clumpy structure within the overall gradients.
In addition, very high resolution observations of CS \jj54\ emission toward
B335
are not consistent with predicted line profiles very close to the forming
star (Wilner et al 1999); either CS is highly depleted in the infalling
gas, or the velocity or density fields depart from the model.

\begin{figure}[ht!]
\centering
\vspace*{11.5cm}
   \leavevmode
   \includegraphics{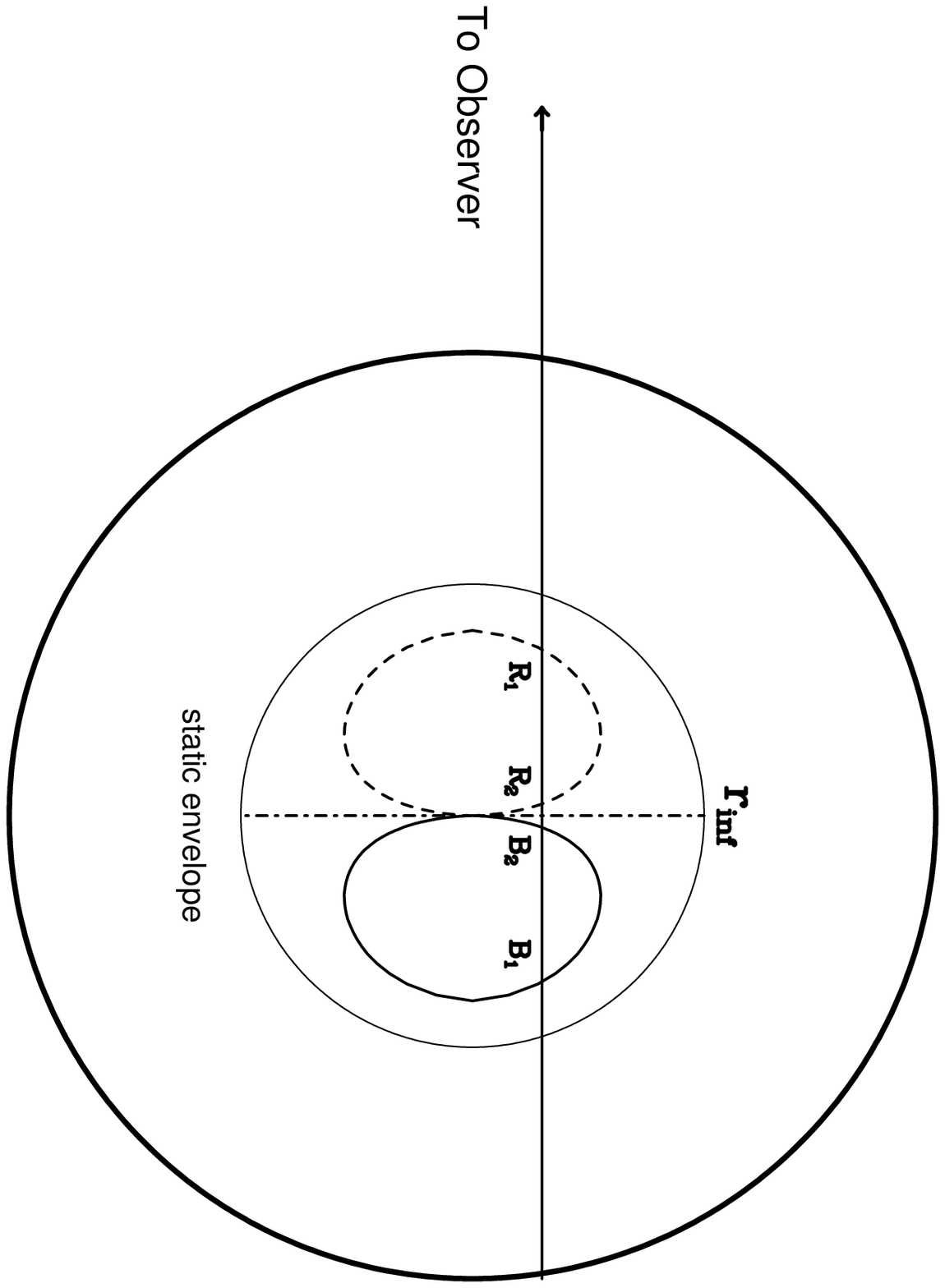}

\caption{ A schematic explanation of why line profiles of optically thick,
high-excitation lines are skewed to the blue in a collapsing cloud.
The ovals are loci of constant {\it line-of-sight} velocity, for
$v(r) \propto r^{-0.5}$. Each line of sight intersects these loci at
two points. The point closer to the center will have a higher \tex,
especially in lines that are hard to excite, so that
$\tex(R_2) > \tex(R_1)$ and $\tex(B_2) > \tex(B_1)$. If the line is
sufficiently
opaque, the point $R_1$ will obscure the brighter $R_2$, but $B_2$ lies
in front of $B_1$. The result is a profile with the blue peak stronger
than the red peak (Zhou \& Evans 1994).
}

\end{figure}

If, for the sake of argument, we accept a blue profile as a collapsing core,
can we see any evolutionary trends? A number of surveys for blue profiles have
been undertaken recently. Gregersen et al (1997) found 9 blue profiles in
23 Class 0 sources, using the \jj32\ and \jj43\ lines of \hcop\ and \hcopi.
After consideration of possible confusion by outflows, etc., they identified
6 sources as good candidates for collapse.  Mardones et al (1997) extended
the search to Class I sources with $\tbol < 200$ K,
using CS and \form\ lines. They introduced the line asymmetry
as a collapse indicator:

\begin{equation}
\delta V = (V_{thick} - V_{thin})/\Delta V_{thin},
\end{equation}
where $V_{thick}$ is the velocity of the peak of the opaque line, $V_{thin}$
is the velocity of the peak of the optically thin line, and $\Delta V_{thin}$
is the linewidth of the thin line.
They confirmed many of the collapse candidates found
by Gregersen and identified 6 more, but they found very few collapse candidates
among the Class I sources. The difference could be caused by using different
tracers, since the CS and \form\ lines are less opaque than the \hcop\ lines.
To remove this uncertainty, Gregersen (1998) surveyed the Class I
sources of Mardones et al (1997) in \hcop. Using $\delta V$ as the measure,
the fraction of blue profiles did not decrease substantially between
Class 0 and Class I. Most of these line profiles need further observations
before they become bona fide candidates.

When does collapse begin? Surveys of the Class $-1$ cores might reveal
very early stages of collapse.  In the inside-out collapse
picture, blue profiles should appear, if at all, only toward the center.
In fact, blue profiles have been seen in a substantial number of these cores
(Gregersen 1998, Lee et al 1999) and maps of one, L1544, show that the
blue profiles are very extended spatially (Tafalla et al 1998).
Clearly, extended blue profiles in Class $-1$ cores do not fit expectations
for early collapse, and Tafalla et al argue that the velocities are higher
than expected for ambipolar diffusion. If the regions producing the
blue and red peaks are indeed
approaching one another, they are forming a core more rapidly than in
standard models, suggesting that some new ideas
may be necessary (e.g. Nakano 1998, Myers \& Lazarian 1998).
For a current review of this field, see Myers et al (2000).

\subsection{Summary of Isolated Star Formation}    \label{summaryiso}

Distinct cores can be identified in tracers of dense ($\n \geq \eten4$
\cmv) gas; these cores are frequently associated with star formation.
There is no clear evidence that they are magnetically subcritical, and some
kinematic evidence suggests that the decay of turbulence, rather than
ambipolar diffusion, is the critical feature. An empirical evolutionary
sequence, based on the spectral appearance of dust emission, and detailed
theoretical models are now being tested by observations. The spatial
distribution of dust emission is providing important tests by probing
$n(r)$. Predictions of the evolution of spectral lines during collapse
are available for the simplest theory and observations of some sources
match predictions of theory quite well. Evidence of collapse is now
strong in a few cases and surveys for distinctive line profiles have
revealed many more possible cases.

\section{CLUSTERED STAR FORMATION AND MASSIVE STARS}  \label{clustered}

In this section, we will address the issues of clustered formation, regardless
of mass, and high-mass star formation, which seems to occur exclusively in
clusters. In this review, the term cluster refers to a group of forming stars,
whether or not the eventual outcome is a bound cluster.
Since high mass stars are rare, the nearest examples are more
distant than is the case for low mass star formation. Together with the
fact that they form in clusters, the greater distance makes it difficult
to isolate individual events of star formation. On the other side of the
balance, massive stars are more easily detectable at large distances because
luminosity is such a strong function of mass. Heating of the surroundings
makes them strong emitters in both dust and many spectral lines; the spectra
of regions forming massive stars are often very rich, reflecting both
a high excitation state and, in many cases, enhanced abundances, as
complex chemistry is driven by elevated temperatures (van Dishoeck \& Blake
1998). These features led early studies, when sensitivity was poor, to
concentrate on regions forming massive stars. However, most of these
advantages arise because the star is strongly influencing its surroundings;
if we wish to know the preconditions, we will be misled. This problem is
aggravated by the fast evolution of massive stars to the main sequence.
Reviews of the topic of clustered star formation can be found in
Elmegreen (1985), Lada \& Lada (1991), and Lada (1999).
Reviews focusing on the formation of massive stars
include Churchwell (1993, 1999), Walmsley (1995), Stahler et al
(2000), and Kurtz et al (2000).

\subsection{Theoretical Issues}                     \label{theorymassive}

Some of the primary theoretical issues regarding the formation of massive stars
have been reviewed by Stahler et al (2000). First, what is the relevant
dividing
line between low-mass and high-mass stars? For the star formation problem, the
question is how far in mass the scenario for low-mass stars can be extended.
Theoretically, the limit is probably about 10 \msun, where stars reach the
main sequence before the surrounding envelope is dissipated. Observations
of the physical conditions in regions forming intermediate mass stars
(Herbig Ae/Be stars and their more embedded precursors) can reveal whether
modifications are needed at even lower masses. Since accretion through
disks plays a crucial role in the standard model, it is important to know
the frequency and properties of disks around more massive stars.

Stars as massive as 100 \msun\ seem to exist (Kudritzki et al 1992), but
radiation pressure from the rapidly evolving stellar core should stop accretion
before such masses can be built (e.g. Wolfire \& Cassinelli 1987).
In addition, massive stars produce very strong outflows (Shepherd \& Churchwell
1996, Bachiller 1996). Since standard accretion theory cannot produce
rates of mass accretion high enough to overwhelm these dispersive effects,
some new effects must become important.

Related questions concern the formation of clusters.
To what extent can the ideas of isolated core
collapse be applied if there are competing centers of collapse nearby?
If the collapse to form massive stars is supercritical,
a whole region may collapse and fragment to form many stars.

The fact that the most massive stars are found near the centers of forming
clusters has led to the suggestion that massive stars are built by collisional
coalescence of stars or protostars (Bonnell et al 1997, 1998). This scheme
requires high densities ($n_{\star} \geq \eten4$ stars pc$^{-3}$); for
a mean stellar mass of 1 \msun, this corresponds to $n \geq 2\ee5$ \cmv. Even
higher stellar densities are seen in the core of the Orion Nebula Cluster
(Hillenbrand \& Hartmann 1998).

The special problems of making the most massive stars are a subset of
the larger question of explaining the mass distribution of all stars.
There may be variations in the IMF between clusters (Scalo 1998), which
can test theories. Hipparcos observations of nearby OB associations have
extended the membership to lower mass stars (de Zeeuw et al 1999), suggesting
total masses of a few \eten3\ \msun.

The main questions these issues raise for observations are whether the mass
and density of cores are sufficient for forming clusters and massive stars,
and whether the mass distribution of clumps in a star forming region can be
related to the mass distribution of stars ultimately formed.

\subsection{Overall Cloud and Core Properties} \label{propmassive}

What do we know about the general properties of the galactic clouds?
The broadest picture is provided by surveys of CO and \coo. Surveys of
significant areas of the Galaxy indicate that the power-law distribution in
mass seen for small clouds ( $dN(M) \propto M^{-\alpha} dM$, \S \ref{probecol})
continues up to a cutoff at $M \sim 6\ee6$ \msun\ (Williams \& McKee 1997).
Studies by Casoli et al (1984),
Solomon \& Rivolo (1989), and Brand \& Wouterloot (1995) find $1.4 \leq
\alpha \leq 1.8$ over both inner and outer Galaxy. Extinction surveys find
flatter slopes (Scalo 1985), but Scalo \& Lazarian (1996) suggest that
cloud overlap affects the extinction surveys.
The fact that $\alpha < 2$ implies that most of the mass is in the largest
structures, although there are issues of how to separate clouds at the
largest scales. Since the star formation rate per unit mass, measured by CO,
appears not to depend on cloud mass (Mead et al 1990, Evans 1991), the mass
distribution supports the idea that most stars form in massive clouds
(Elmegreen 1985). However, the enormous spread ($> \eten2$) in star formation
rate per unit mass at any given mass, together with the fact that most of
the molecular gas is sterile, suggests that comparisons to overall cloud
masses, measured by CO, are not particularly relevant.

Surveys in molecular lines indicative of denser gas have generally been biased
towards signposts of star formation.  Exceptions are the CS \jj21\ survey of
L1630 (Lada et al 1991) and the CS \jj10\ and \jj21\ surveys of L1641
(Tatematsu et al 1993, 1998). These two clouds, also called Orion B and
Orion A, are adjacent. In both cases, the CS \jj21\ maps showed more contrast
than maps of \coo\ (Bally et al 1987), with CS \jj10\ being somewhat
intermediate. Maps of higher-$J$ transitions have been less complete,
but show still less area covered by emission.
The CS surveys detected less than 20\% of the total mass in
both clouds (Lada et al 1991, Tatematsu, personal communication).
The \jj21\ emission in L1641 is somewhat smoother than the \jj21\ emission
from L1630 (Tatematsu et al 1998), and this difference may be reflected in the
distribution of star formation.
Star formation in L1641 appears to include a distributed component (Strom et al
1993); in contrast, star formation in L1630 is tightly concentrated in clusters
associated with massive cores of dense gas (Lada 1992, Li et al 1997).

Excitation analysis of CS lines with higher critical density in L1630 shows
that the
star-forming regions all contain gas with $n \geq \eten5$ \cmv\
(Lada et al 1997).
These results suggest that surveys in lines of high \nc\ are relevant for
characterizing star formation regions. Of the possible tracers, CS and
\ammonia\ have been most widely surveyed.
In comparison to cores in Taurus, where only low-mass stars are forming, cores
in the Orion clouds tend to be more massive and to have larger linewidths
when observed with the same tracer (CS: Tatematsu et al 1993, \ammonia: Harju
et al 1993). The differences are factors of
2--4 for the majority of the cores, but larger for the cores near the Orion
Nebula and those forming clusters in L1630 (Lada et al 1991).

CS transitions have been surveyed toward
Ultra-Compact (UC) HII regions, \water\ masers,
or luminous \iras\ sources (see Kurtz et al 2000). Since the \iras\ survey
became available, most samples are drawn from the \iras\ catalog
with various color selection criteria applied.
The most complete survey (Bronfman et al 1996) was toward \iras\ sources
with colors characteristic of UC HII regions (Wood and Churchwell 1989)
over the entire Galactic plane.
Bronfman et al found CS \jj21\ emission (see Table 1 for density
sensitivity) in 59\% of 1427 \iras\ sources,
and the undetected sources were either weak in the \fir\ or had
peculiar colors.
Searches toward \water\ masers have used the catalogs of
Braz and Epchtein (1983) and Cesaroni et al (1988).
Surveys of CS \jj21\ (Zinchenko et al 1995, Juvela 1996)
toward southern \water\ masers found detection rates close to 100\%,
suggesting that dense, thermally excited gas surrounds the compact,
ultra-dense regions needed to produce \water\ masers.
The detection rate drops in higher $J$ transitions of CS (Plume et al
1992, 1997) but is still 58\% in the CS \jj76\ line, which probes
higher densities (Table 1).
An LVG, multitransition study of CS lines
found $\mean{{\rm log}n (\cmv)} = 5.9$ for 71 sources and a similar
result from a smaller sample using C$^{34}$S data (Plume et al 1997).
Densities derived assuming LVG fall between
the average and maximum densities in clumpy models with a range of densities
(Juvela 1997, 1998).

Maps of the cores provide size and mass information. Based on the
sizes and masses of 28 cores mapped in the CS \jj21\ line (Juvela 1996),
one can compute a mean size, $\meanl = 1.2 \pm 0.5$ pc,  mean
virial mass, $\mean{\mv} \sim 5500$ \msun, and $\mean{\mc} \sim 4900$
\msun.  While the two mass estimates agree on average, there can be large
differences in individual cases.
Remarkably, cloud structure does not introduce a big uncertainty
into the cloud masses: using a clumpy cloud model, (see \S \ref{clumps}),
Juvela (1998) found that $\mc$ increased by a factor of 2 on average compared
to homogeneous models and agreed with $\mv$ to within a factor of 2.
Plume et al (1997) obtained similar results from strip maps of CS \jj54:
$\meanl = 1.0 \pm 0.7$ pc (average over 25 cores); $\mean{\mv} = 3800$
\msun\ (16 cores). As usual mean values must be regarded with caution;
there is a size distribution. As cores with weaker emission are mapped,
the mean size decreases; an average of 30 cores with full maps of \jj54\
emission gives $\meanl = 0.74 \pm 0.56$ pc, with a range of 0.2 to 2.8 pc
(Y Shirley, unpublished results).

Churchwell et al (1992) surveyed 11 UC HII regions for CS \jj21\ and
\jj54\ emission, leading to estimates of $n \geq \eten 5$ \cmv.
Cesaroni et al (1991) surveyed 8 UC HII in three transitions of CS and
C$^{34}$S and estimated typical sizes of 0.4 pc, masses of 2000 \msun,
and densities of \eten 6 \cmv.
More extensive surveys have been made in \ammonia; Churchwell et al (1990)
found \ammonia\ $(J,K) = (1,1)$ and (2,2) emission from 70\% of a sample
of 84 UC HII regions and \iras\ sources with similar colors. They derived
\tk, finding a peak in the distribution around 20 K, but a significant
tail to higher values. Further studies in the (4,4) and (5,5) lines toward
16 UC HII regions with strong (2,2) emission (Cesaroni et al 1992)
detected a high fraction.
Estimates for \tk\ ranged from 64 to 136 K and sizes of 0.5 pc. Two sources
indicated much higher densities and \ammonia\ abundances. Follow-up studies
with the VLA  (Cesaroni et al 1994, 1998)
revealed small ($\sim 0.1$ pc), hot ($\tk \sim 50 -200$ K),
dense ($n = \eten 7$ \cmv ) regions with enhanced \ammonia\ abundances.
These hot cores (discussed below) are slightly displaced from the UC HII,
but coincide with \water\ masers.

Magnetic fields strengths have been measured with the Zeeman effect toward
about 10 regions of massive star formation. The fields are substantially
stronger than the fields seen in isolated, low-mass cores, but the masses
are also much higher. In most cases the mass to flux ratio is comparable
to the critical ratio, once geometrical effects are considered
(Crutcher 1999b). Given the uncertainties and sample size, it is too early
to decide if regions forming massive stars are more likely to be supercritical
than regions forming only low-mass stars.

The ionization fraction in the somewhat more massive Orion cores appears
to be very similar to that in low-mass cores: $-6.9 < \log x_e < -7.3$
(Bergin et al 1999).  The most massive cores in their sample have
$x_e \leq \eten{-8}$, as do some of the massive cores studied by de
Boisanger et al (1996). Expressed in terms of column density, the decline
in $x_e$ appears around $ N \sim 3\ee{22}$ \cmc. At $x_e = \eten{-8}$,
$t_{AD} = 7\ee5$ yr, about $0.1 t_{AD}$ in isolated, low-mass cores.
Even if the massive cores are subcritical, their
evolution should be faster than that of low-mass cores.

To summarize, the existing surveys show ample evidence that massive star
formation usually takes place in massive ($M > \eten3$ \msun ), dense
($n \sim \eten6$ \cmv ) cores, consistent with the requirements inferred
from the study of young clusters and associations, and with conditions needed
to form the most massive stars by mergers. Cores with measured $B_z$
seem to be near the boundary between subcritical and supercritical cores.

\subsection{Evolutionary Scenarios and Detailed Theories}
\label{evolmassive}

To what extent can an evolutionary scenario analogous to the class system
be constructed for massive star formation? Explicit attempts to fit
massive cores into the class system have relied on surveys of \iras\ sources.
Candidates for massive Class 0 objects, with $L>\eten3$ \lsun, have been found
(Wilner et al 1995, Molinari et al 1998).
One difficulty with using the shape of the spectral energy distribution
for massive star formation is that dense, obscuring material usually
surrounds objects even after they have formed, and a single star may be quite
evolved  but still have enough dust  in the vicinity to have the same
spectral energy distribution as a much younger object. The basic
problem is the difficulty in isolating single objects.  Also, the role of disks
in massive regions is less clear, and they are unlikely to
dominate the spectrum, as they do in low mass Class II sources.
Other markers, such as the detection of radio continuum emission,
must be used as age indicators. Hot cores provide
obvious candidiates to be precursors of UC HII regions, but some have
embedded UC HII regions and may be transitional (Kurtz et al 2000).
The chemical state of massive cores may also provide an evolutionary
sequence; Helmich et al (1994) suggested an evolutionary ordering of
three sources in the W3 region based on their molecular spectra and
chemical models (see van Dishoeck \& Blake 1998).

While theories for clustered and massive star formation are much less
developed, some steps have been taken (e.g. Bonnell et al 1998, Myers 1998).
The larger \dv\ in regions forming
massive stars imply that turbulence must be incorporated into the models.
Myers \& Fuller (1992) suggested a ``thermal-non-thermal" (TNT) model,
in which $\n(r)$ is represented by the sum of two power-laws and the
term with $p = 1$ dominates outside the radius ($r_{TNT}$) at which turbulent
motions dominate thermal motions. McLaughlin \& Pudritz (1997) develop
the theory of a logatropic sphere, which has a density distribution
approximated by $p = 1$. Collapse in such a configuration leads to
power laws in the collapsing region with similar form to those in the
collapsing isothermal sphere, but with
higher densities and lower velocities. These ideas lead to accretion rates
that increase with time and timescales for
forming massive stars that are much weaker functions of the final stellar mass
than is the case for the isothermal sphere (\S \ref{theorydet}).
Recent simulations of unmagnetized fragmentation that follow the interaction
of clumps find that the mass spectrum of fragments steepens from $\alpha = 1.5$
to a lognormal distribution of the objects likely to form stars (e.g. Klessen
et al 1998). To avoid an excessive global star formation rate (\S \ref{defs})
and distortion of the clump mass spectrum in the bulk of the cloud,
this process must be confined to star-forming regions in clouds.

\subsection{Filaments, Clumps, Gradients, and Disks}           \label{clumps}

An important issue is whether the dense cores have overall density
gradients or internal structures (clumps) that are likely to form
individual stars and whether the mass distribution is like that of stars.
Unfortunately, the terms ``clumps" and ``cores" have no standard usage; I
will generally use ``cores" to refer to regions that appear in maps of
high-excitation lines and ``clumps" to mean structures inside cores,
except where a different usage is well established (e.g. ``hot cores").
Myers (1998) has suggested the term ``kernels" to describe clumps within cores.
Cores themselves are usually embedded in structures traced by lines
with lower critical density, and these are also called clumps by those who map
in these lines. Many of these low-density structures are quite filamentary
in appearance: examples include the \coo\ maps of L1641 (Orion B) of
Bally et al (1987). In some cases, this filamentary structure is seen
on smaller scales in tracers of high density or column density (e.g.
Johnstone \& Bally 1999, Figure 2).

The clumpy structure of molecular clouds measured in low-excitation
lines suggests that dense cores will be clumpy as well. Suggestions
of clumpy structure came from early work comparing densities derived
from excitation analysis in different tracers, but smooth density gradients
provided an alternative (e.g. Evans 1980). Multitransition studies
of three cores
in CS (Snell et al 1984), C$^{34}$S (Mundy et al 1986) and \form\
(Mundy et al 1987) found no evidence for overall density gradients; the
same high densities were derived over the face of the core, while the strength
of the emission varied substantially. This was explained in a clumpy model
with clump filling factors of the dense gas $f_v \sim 0.03$ to 0.3, based on a
comparison of \mv\ with \mn\ (Snell et al 1984).
This comparison forms the basis for most claims of unresolved clumps.

Observations with higher resolution support the idea of clumps postulated
by Snell et al (1984). For example, Stutzki and G\"usten (1990)
deconvolved 179 clumps from a map of \cooo\ \jj21\ emission near M17.
Because of overlap, far fewer clumps are apparent to the eye; assumptions
about the clump shape and structure may affect the deconvolution.
Maps of the same source in several CS and C$^{34}$S lines
(Wang et al 1993) could be reproduced with the clump catalog of Stutzki
\& G\"usten, but only with densities about 5 times higher than they found.
Thus the clumps themselves must have structure, either a continuation
of clumpiness or smooth gradients. Since the inferred clumps are now
similar in size to the cores forming low mass stars, a natural question
is whether massive cores are fragmented into many clumps which can
be modeled as if they were isolated cores. In favor of this view,
Stutzki and G\"usten noted that the Jeans length was similar to the
size of their clumps.

A significant constraint on this picture is provided by the close
confinement of the clumps; unlike the picture of isolated core formation,
the sphere of influence of each clump will be limited by its neighbors.
A striking example is provided by the dust continuum maps of the
$\rho$ Ophiuchi cloud (Figure 1), our nearest example of cluster formation,
albeit with no very massive stars. Within about six cores of size 0.2 pc,
Motte et al (1998) find about 100 structures with sizes of 1000-4000 AU.
They deduce a fragmentation scale of 6000 AU, five times smaller
than isolated cores in Taurus. Thus the ``feeding zone" of an individual
clump is considerably less and the evolution must be more dynamic,
with frequent clump-clump interactions, than is the case for isolated
star formation.
This picture probably applies even more strongly to the
more massive cores. In $\rho$ Ophiuchi, the clump mass spectrum above 0.5
\msun\
steepens to $\alpha = 2.5$, close to the value for stars (Motte et al 1998),
in agreement with predictions of Klessen et al (1998).
A similar result ($\alpha = 2.1$)
is found in Serpens, using \mm\ interferometry (Testi \& Sargent 1998).
While more such studies are needed,
these results are suggesting that dust continuum maps do
trace structures that are likely precursors of stars, opening the study
of the origin of the IMF to direct observational study.

Some of the less massive, relatively isolated cores, such as NGC2071,
S140, and GL2591, have been modeled with smooth density
and temperature gradients (Zhou et al 1991, 1994b, Carr et al 1995,
van der Tak et al 1999).
Models with gradients can match the relative strengths of a range
of transitions with different excitation requirements, improving on
homogeneous models.
Zhou et al (1994) summarized attempts to deduce gradients and found
preliminary evidence that, as core mass increases, the tendency is first
toward smaller values of $p$. The most massive cores showed little evidence
for any overall gradient and more tendency toward clumpy substructure.
This trend needs further testing, but it is sensible if more massive cores
form clusters. Lada et al (1997) found that the L1630 cores forming rich
embedded clusters with high efficiency tended to have larger masses of dense
($n > \eten5$ \cmv) gas, but a lower volume filling factor of such gas,
indicating more fragmentation.

However, the line profiles of optically thick lines
predicted by models with gradients are usually self-absorbed, while
the observations rarely show this feature in massive cores. Clumps
within the overall gradients are a likely solution.
The current state of the art in modeling line profiles in massive cores
is the work of Juvela (1997, 1998),
who has constructed clumpy clouds from both structure tree and fractal
models, performed 3-D radiative transport and excitation,
and compared the model line profiles
to observations of multiple CS and C$^{34}$S transitions in
massive cores. He finds that the clumpy models match the line profiles
much better than non-clumpy models, especially if macroturbulence dominates
microturbulence. Structure trees (Houlahan \& Scalo 1992) match the data
better than fractal models, but overall density and/or temperature gradients
with $p + q \approx 2$ are needed in addition to clumps.

The study of gradients versus clumps in regions forming intermediate mass
stars could help to determine whether conditions change qualitatively
for star formation above some particular mass and how this change is related
to the outcome. Using \nir\ observations of regions with Herbig Ae/Be
stars, Testi et al (1997) found that the cluster mode of star
formation becomes dominant when the most massive star has a spectral type
earlier than B7. Studies of the \fir\ emission
from dust remaining in envelopes around Herbig Ae/Be stars found values
of $p$ ranging from 0.5 to 2 (Natta et al 1993). Maps of dust continuum
emission
illustrate the difficulties: the emission may not peak on the visible star,
but on nearby, more embedded objects (Henning et al 1998, Di Francesco et al
1998).
Detailed models of several suitable sources yield $p = 0.75$ to 1.5 (Henning
et al 1998, Colom\'e et al 1996). Further work is needed to determine whether
a change in physical conditions can be tied to the change to cluster mode.

Many Herbig Ae stars have direct evidence for disks from interferometric
studies
of dust emission (Mannings \& Sargent 1997), though fewer Herbig Be stars have
such direct evidence (Di Francesco et al 1997).
For a review, see Natta et al (2000).
Disks may be more common during more embedded phases of B star formation;
Shepherd and Kurtz (1999) have found a large (1000 AU) disk around an
embedded B2 star.
The statistics of UC HII regions may provide indirect evidence
for disks around more massive stars. Since such regions should expand
rapidly unless confined, the large number of such regions posed a puzzle
(Wood \& Churchwell 1989).
Photoevaporating disks have been suggested as a solution
(Hollenbach et al 1994).
Such disks have also been used to explain
very broad recombination lines (Jaffe \& Mart\'{\i}n-Pintado 1999).
Kinematic evidence for disks will be discussed in \S \ref{kinmassive}.

A particular group of clumps (or cores) deserve special mention: hot cores
(e.g. Ohishi 1997). First identified in the Orion cloud
(Genzel \& Stutzki 1989),
about 20 are now known (see Kurtz et al 2000). They are small regions
($l \sim 0.1$ pc), characterized by $\tk > 100$ K, $n > \eten7$ \cmv,
and rich spectra, probably reflecting enhanced abundances, as well as high
excitation (van Dishoeck \& Blake 1998). Theoretical issues have been reviewed
by Millar (1997) and Kaufman et al (1998), who argue that they are likely to
be heated internally, but they often lack radio continuum emission. Since
dynamical timescales for gas at such densities are short, these may plausibly
be precursors to the UC HII regions.

The evidence for flatter density distribution in regions of intermediate
mass support the relevance of models like the TNT or logatropic sphere models
in massive regions (\S \ref{evolmassive}), but it will be important to
study this trend with the same methods now being applied to regions forming
low mass stars, with due regard for the greater distance to most regions
forming massive stars. The increasingly fragmented structure in
more massive cores and the increased frequency of clusters above a certain
mass are consistent with a switch to a qualitatively different mode
of star formation, for which different theories are needed. Finally,
the common appearance of filaments may support a continuing role for
turbulence in dense regions, since simulations of turbulence often
produce filamentary structure (e.g. Scalo et al 1998).

\subsection{Kinematics}                           \label{kinmassive}

Lada et al (1991) found only a weak correlation between linewidth
and size, disappearing entirely for a different clump definition,
in the L1630 cores (Goodman Type 2 relation, see \S \ref{internal}). Caselli
\& Myers (1995) also found that the Type 1 linewidth-size relation (with only
non-thermal motions included, $\Delta v_{NT} \propto R^{\gamma}$) is
flatter in massive cloud cores ($\gamma = 0.21\pm 0.03$) than in low mass
cores ($\gamma = 0.53 \pm 0.07$); in addition, the correlation is poor
(correlation coefficient of 0.56) though the correlation is better for
individual cores (Type 3 relations). They also noted that non-thermal
(turbulent) motions are much more dominant in more massive cores and find
good agreement with predictions of the TNT model. The
``massive" cores in the Caselli \& Myers study are mostly the cores in
Orion with masses between 10 and 100 \msun. The much more massive
($\mean{\mv} = 3800$ \msun) cores studied by Plume et al (1997) exhibit
no statistically significant linewidth-size relation at all (correlation
coefficient  is 0.26) and the linewidths are systematically higher (by
factors of 4--5) for a given size than would be predicted by the
relationships derived for low and intermediate mass cores (Caselli \& Myers
1995).
In addition, the densities of these cores exceed by factors of 100 the
predictions of density-size relations found for less massive cores (Myers
1985). The regions forming truly massive stars are much more dynamic, as
well as much denser, than would be expected from scaling relations found
in less massive regions. The typical linewidth in massive cores is
6--8 \kms, corresponding to a 1-D velocity dispersion of 2.5--3.4 \kms,
similar to that of the stars in the Orion Nebula Cluster (Hillenbrand
\& Hartmann 1998). Larson (1981) noted that regions of massive star
formation, like Orion and M17, did not follow his original linewidth-size
relation, suggesting that gravitational contraction would decrease size
while keeping \dv\ roughly constant or increasing it.

Searching for collapse in massive cores is complicated by the turbulent,
clumpy structure, with many possible centers of collapse, outflow, etc.
A collapse signature may indicate an overall collapse of the core, with
accompanying fragmentation. In fact, self-absorbed line profiles from
regions forming massive stars are rather rare (Plume et al 1997).
A possible collapse signature has been seen in CS emission toward NGC~2264 IRS,
a Class I source with $L \sim 2000$ \lsun\ (Wolf-Chase \& Gregersen 1997).
If an HII region lies at the center of a
collapsing core, absorption lines should trace only the gas in front
and should be redshifted relative to the emission lines. Failure to see
this effect, comparing \form\ absorption to CO emission, supported an
early argument against the idea that all clouds were collapsing (Zuckerman
\& Evans 1974). A more recent application of this technique to dense cores,
using high-excitation lines of \ammonia, showed no preference for red-shifted
absorption (Olmi et al 1993) overall, but a few dense cores do show this kind
of effect. These sources include W49 (Welch et al 1988, Dickel \& Auer 1994),
G10.6--0.4 (Keto et al 1988), and W51 (Zhang \& Ho 1997, Zhang et al 1998a).

Dickel \& Auer (1994) tested different collapse scenarios against observations
of \hcop\ and favored free-fall
collapse, with $n \propto r^{-1.5}$ and $v \propto r^{-0.5}$ throughout
W49A North; they noted that more complex motions are present on small scales.
Keto et al (1988) used \ammonia\ observations with 0.3\as\ resolution to
separate infall from rotational motions toward the UC HII region, G10.6--0.4.
Zhang \& Ho (1997) used \ammonia\ absorption and Zhang et al (1998a) added
CS \jj32\ and CH$_3$CN observations to identify collapse onto two UC HII
regions, W51e2 and W51e8, inferring infall velocities of about 3.5 \kms\ on
scales of 0.06 pc. Young et al (1998) have tested various collapse models
against
the data on W51e2 and favor a nearly constant collapse velocity ($v \sim 5$
\kms) and $n(r) \propto r^{-2}$.
Mass infall rates of about 6\ee{-3} (Zhang et al 1998a) to 5\ee{-2} (Young et
al 1998) \msun\ yr$^{-1}$ were inferred for W51e2.
Similar results were found in G10.6--0.4 (Keto et al 1988), and even more
extreme mass infall rates (\eten{-2} to 1 \msun\ yr$^{-1}$) have
been suggested for W49A, a distant source with enormous mass
($\sim \eten6 \msun$) and luminosity ($L \sim \eten7$ \lsun) (Welch et al
1988). These high infall rates may facilitate formation of very massive stars
(\S \ref{theorymassive}) and help confine UC HII regions (Walmsley 1995).

Large transverse velocity gradients have been seen in some hot cores, including
G10.6--0.4 (Keto et al 1988), W51 (Zhang \& Ho 1997), and $IRAS 20126+4104$
(Cesaroni et al 1997, Zhang et al 1998b). For example,
gradients reach 80 \kms\ pc$^{-1}$ in the \ammonia\ $(J,K) = (4,4)$ line
and 400 \kms pc$^{-1}$ in the CH$_3$CN \jj65 line toward G29.96--0.02
and G31.41+0.31 (Cesaroni et al 1994b, 1998). Cesaroni et al (1998)
interpret the gradients in terms of rotating disks extending to about
\eten4 AU. Using the dust continuum emission
at 3 mm, they deduce disk masses up to 4200 \msun, in the case of
G31.41+0.31.

\subsection{Implications for Larger Scales}               \label{implications}

Since most stars form in the massive cores discussed in this section (Elmegreen
1985), they are most relevant to issues of star formation on a galactic scale.
If the luminosity is used to trace star formation rate (e.g. Rowan-Robinson
et al 1997, Kennicutt 1998), the star formation
rate per unit mass is proportional to $L/M$. Considering clouds as a whole,
$\mean{L/M} \sim 4$ in solar units (e.g. Mooney \& Solomon 1988),
with a spread exceeding
a factor of $\eten2$ (Evans 1991). Using CS \jj54\ emission to measure $M$
in the dense cores, Plume et al (1997) found $\mean{L/M} = 190$ with a
spread of a
factor of 15. The star formation rate per unit mass is much higher and less
variable if one avoids confusion by the sterile gas. The average $L/M$
seen in dense cores in our Galaxy is similar to the highest
value seen in ultra-luminous infrared galaxies, where $M$ is measured by CO
(Sanders et al 1991, Sanders \& Mirabel 1998).
The most dramatic starburst galaxies behave as if their entire
interstellar medium has conditions like those in the most active, massive
dense cores in our Galaxy.
Some studies (e.g. Mauersberger \& Henkel 1989) indeed found strongly
enhanced CS \jj54\ emission from starburst galaxies.
It will be interesting to observe high-$J$ CS lines in the most
luminous galaxies to compare to the conditions in massive cores in our Galaxy.
Perhaps the large scatter in $L/M$ seen in galaxies
will be reduced if CS, rather than CO, is used as a measure of the gas,
ultimately leading to a better understanding of what controls galactic
star formation rates (see Kennicutt 1998).

\subsection{Summary of Clustered Star Formation}    \label{summarymassive}

The cloud mass distribution found for lower mass objects continues to
massive clouds, but less is known about the distribution for dense cores.
Very massive cores clearly exist, with sufficient mass ($M > \eten3$ \msun)
to make the most massive clusters and associations. These cores are denser
and much more dynamic than cores involved in isolated star formation, with
typical $n \sim \eten6$ \cmv\ and linewidths about 4--5 times larger than
predicted
from the linewidth-size relation. Pressures in massive cores (both thermal
and turbulent) are substantially higher than in lower mass cores.
The densities match those needed to
form the densest clusters and the most massive stars by coalescence.
There are some cores with evidence of overall collapse, but most do not
show a clear pattern. There is some evidence that
more massive regions have flatter density profiles, and that fragmentation
increases with mass, but more studies are needed.
High resolution studies of nearby regions of cluster formation are finding
many clumps, limiting the feeding zone of a particular star-forming event to
$l \sim 6000$ AU, much smaller than the reservoirs available in the isolated
mode. In some cases, the clump mass distribution approaches
the slope of the IMF, suggesting that the units of star formation have been
identified. Studies of intermediate mass stars indicate that a transition
to clustered mode occurs at least by a spectral type of B7.




\section{CONCLUSIONS AND FUTURE PROSPECTS}

The probes of physical conditions have been developed and are now fairly
well understood. Some physical conditions have been hard to measure, such
as the magnetic field strength, or hard to understand, notably the kinematics.
Star-forming structures, or cores, within primarily sterile molecular clouds
can be identified by thresholds in column density or density. Stars form
in distinct modes, isolated and clustered, with massive stars forming
almost exclusively in a clustered mode. The limited number of measurements
of magnetic field leave open the question of whether cores are subcritical
or supercritical, and whether this differs between the isolated and clustered
mode. Cores involved in isolated star formation may be distinguished from
their surroundings by a decrease in turbulence to subsonic levels, but
clustered
star formation occurs in regions of enhanced turbulence and higher density,
compared to isolated star formation.

Evolutionary scenarios and detailed theories exist for the isolated mode.
The theories assume cores with extended, power-law density gradients, leading
to the mass accretion rate as the fundamental parameter. Detailed tests
of these ideas are providing overall support for the picture, but also
raising questions about the detailed models. Notably, kinematic evidence
of gravitational collapse has finally been identified in a few cases.
The roles of turbulence, the magnetic field, and rotation must be understood,
and the factors that bifurcate the process into single or multiple star
formation must be identified.
Prospects for the future include a less biased census for cores in early
stages and improved information on density gradients, both facilitated by
the appearance of cameras at \mm\ and \submm\ wavelengths on large telescopes.
Antenna arrays operating at these wavelengths will provide more detailed
information on the transition region between envelope and disk and study
early disk evolution and binary fraction. Future, larger, antenna arrays will
probe disk structure to scales of a few AU. Finally, a closer coupling of
physical and chemical studies with theoretical models will provide more pointed
tests of theory.

Theories and evolutionary scenarios are less developed for the clustered
mode, and our understanding of the transition between isolated and clustered
mode is still primitive. Current knowledge suggests that more massive cores
have flatter density distributions and greater tendency to show substructure.
At some point, the substructure dominates and multiple centers of collapse
develop. With restricted feeding zones, the mass accretion rate gives way
to the mass of clump as the controlling parameter, and some studies of clump
mass spectra suggest that the stellar IMF is emerging in the clump mass
distribution.  The most massive
stars form in very turbulent regions of very high density. The masses and
densities are sufficient to form the most massive clusters and to explain
the high stellar density at the centers of young clusters.
They are also high enough
to match the needs of coalescence theories for the formation of the most
massive
stars. As with isolated regions, more and better measurements of magnetic
fields
are needed, along with a less biased census, particularly for cool cores that
might represent earlier stages. Larger antenna arrays will be able to separate
clumps in distant cores and determine mass distributions for comparison to the
IMF.  Larger airborne telescopes will provide complementary information
on luminosity sources in crowded regions. Observations with
high spatial resolution and sensitivity in the \mir\ will provide
clearer pictures of the deeply embedded populations,
and \mir\ spectroscopy with high spectral resolution could trace kinematics
close to the forming star.  Deeper understanding of clustered star formation
in our Galaxy will provide a foundation for understanding the origin and
evolution of galaxies.

ACKNOWLEDGEMENTS

I am grateful to P Andr\'e, J Bally, M Choi, F Motte, D Johnstone, and L
Looney for supplying figures.
Many colleagues sent papers in advance of publication
and/or allowed me to discuss results in press or in progress. A partial list
includes P Andr\'e, R Cesaroni, R Crutcher, D Jaffe, L Mundy, E Ostriker,
Y Shirley, J Stone, D Ward-Thompson, and D Wilner. I would like to thank
R Cesaroni, Z Li, P Myers, F Shu, F van der Tak, and M Walmsley for detailed,
helpful comments on an earlier version.
This work has been supported by the State of Texas and NASA,
through grants NAG5-7203 and NAG5-3348.


\begin{table}[h]
\caption{Properties of Density Probes}
\vspace {3mm}
\begin{tabular}{l l r r r r r r}
Molecule & Transition & $\nu$ & $E_{up}$ &$n_c (10K)$ & $n_{eff} (10K)$ & $n_c
(100K)$ & $n_{eff} (100K)$ \cr
 &   & (GHz) & (K)& (\cmv) &(\cmv) & (\cmv)  & (\cmv) \\
\cr
CS       & \jj10  & 49.0  & 2.4 & 4.6\ee4 & 7.0\ee3 & 6.2\ee4 & 2.2\ee3 \cr
CS       & \jj21  & 98.0  & 7.1 & 3.0\ee5 & 1.8\ee4 & 3.9\ee5 & 4.1\ee3 \cr
CS       & \jj32  & 147.0 & 14  & 1.3\ee6 & 7.0\ee4 & 1.4\ee6 & 1.0\ee4 \cr
CS       & \jj54  & 244.9 & 35  & 8.8\ee6 & 2.2\ee6 & 6.9\ee6 & 6.0\ee4 \cr
CS       & \jj76  & 342.9 & 66  & 2.8\ee7 & \ldots  & 2.0\ee7 & 2.6\ee5 \cr
CS     & \jj{10}9 & 489.8 &129  & 1.2\ee8 & \ldots  & 6.2\ee7 & 1.7\ee6 \cr
\hcop    & \jj10  & 89.2  & 4.3 & 1.7\ee5 & 2.4\ee3 & 1.9\ee5 & 5.6\ee2 \cr
\hcop    & \jj32  & 267.6 & 26  & 4.2\ee6 & 6.3\ee4 & 3.3\ee6 & 3.6\ee3 \cr
\hcop    & \jj43  & 356.7 & 43  & 9.7\ee6 & 5.0\ee5 & 7.8\ee6 & 1.0\ee4 \cr
HCN      & \jj10  & 88.6  & 4.3 & 2.6\ee6 & 2.9\ee4 & 4.5\ee6 & 5.1\ee3 \cr
HCN      & \jj32  & 265.9 & 26  & 7.8\ee7 & 7.0\ee5 & 6.8\ee7 & 3.6\ee4 \cr
HCN      & \jj43  & 354.5 & 43  & 1.5\ee8 & 6.0\ee6 & 1.6\ee8 & 1.0\ee5 \cr
\form & \fs212111 & 140.8 & 6.8 & 1.1\ee6 & 6.0\ee4 & 1.6\ee6 & 1.5\ee4 \cr
\form & \fs313212 & 211.2 & 17  & 5.6\ee6 & 3.2\ee5 & 6.0\ee6 & 4.0\ee4 \cr
\form & \fs414313 & 281.5 & 30  & 9.7\ee6 & 2.2\ee6 & 1.2\ee7 & 1.0\ee5 \cr
\form & \fs515414 & 351.8 & 47  & 2.6\ee7 & \ldots  & 2.5\ee7 & 2.0\ee5 \cr
NH$_3$ & (1,1)inv & 23.7  & 1.1 & 1.8\ee3 & 1.2\ee3 & 2.1\ee3 & 7.0\ee2 \cr
NH$_3$ & (2,2)inv & 23.7  & 42  & 2.1\ee3 & 3.6\ee4 & 2.1\ee3 & 4.3\ee2 \cr
\cr
\end{tabular}

\end{table}

\clearpage

\newcommand{\apj}{{\it Ap. J.}}
\newcommand{\apjl}{{\it Ap. J. Lett.}}
\newcommand{\apjs}{{\it Ap. J. Suppl.}}
\newcommand{\aj}{{\it A. J.}}
\newcommand{\araa}{{\it Annu. Rev. Astron. Astrophys.}}
\newcommand{\aap}{{\it Astron. Astrophys.}}
\newcommand{\aaps}{{\it Astron. Astrophys. Suppl.}}
\newcommand{\baas}{{\it Bull. Amer. Astro. Soc.}}
\newcommand{\mnras}{{\it MNRAS}}
\newcommand{\qjras}{{\it QJRAS}}
\newcommand{\pasp}{{\it PASP}}
\newcommand{\sci}{{\it Science}}
\newcommand{\nature}{{\it Nature}}
\newcommand{\orlife}{{\it Ori. Life Evol. Biosph.}}
\newcommand{\disksoutflows}{In {\it Disks and Outflows from Young Stars} ed.
SVW Beckwith, J Staude, A Quetz, A Natta Berlin: Springer}
\newcommand{\mcd}{In {\it Frontiers of Stellar Evolution}, ed. DL Lambert
San Francisco:ASP}
\newcommand{\cretei}{In {\it The Physics of Star Formation and Early
Stellar Evolution }, ed. CJ Lada, ND Kylafis. Dordrecht:Kluwer 1991}
\newcommand{\creteii}{In {\it The Physics of Star Formation and Early
Stellar Evolution II}, ed. CJ Lada, ND Kylafis. Dordrecht:Kluwer in press}
\newcommand{\ppi}{In {\it Protostars and Planets}, ed. T Gehrels.
Tucson:Univ. of Arizona}
\newcommand{\ppii}{In {\it Protostars and Planets II}, ed. DC Black, MS
Matthews.
Tucson:Univ. of Arizona}
\newcommand{\ppiii}{In {\it Protostars and Planets III}, ed. EH Levy, JI
Lunine.
Tucson:Univ. of Arizona}
\newcommand{\ppiv}{In {\it Protostars and Planets IV}, ed. V Mannings, A Boss,
S Russell.  Tucson:Univ. of Arizona in press}
\newcommand{\iautokyo}{In {\it Star Forming Regions, IAU Symposium 115}, ed. M
Peimbert J Jugaku. Dordrecht:Reidel}
\newcommand{\iauleiden}{In {\it Molecules in Astrophysics: Probes and
Processes,
 IAU Symposium 178}, ed. EF van Dishoeck. Dordrecht:Reidel}
\newcommand{\haystack}{In {\it Clouds, Cores, and Low Mass Stars}, ed. DP
Clemens, R. Barvainis. San Francisco: ASP}
\newcommand{\cozumel}{Rev. Mex. Astr. Astrof. Conf. Ser. 1}
\newcommand{\isosf}{In {\it Star Formation with the Infrared Space Observatory
(ISO)}, ed. JL Yun, R. Liseau. San Francisco: ASP}
\newcommand{\fesc}{In {\it The Formation and Evolution of Star Clusters}
ed. K Janes. San Francisco:ASP}
\newcommand{\isturb}{In {\it Interstellar Turbulence, Proceedings of the 2nd
Guillermo Haro Conference} ed. J Franco, A Carraminana. Cambridge: Cambridge
Univ. Press in press}
\newcommand{\msism}{In {\it Massive Stars: Their Lives in the Interstellar
Medium} ed. JP Cassinelli, EB Churchwell. San Francisco:ASP}

\setcounter{figure}{0}
\begin{figure}[ht!]

\caption{ (NOTE: this figure is not supplied here. It can be seen in the
original paper (ref. below) and it exists on astro-ph (9903284) as a
gzipped postscript file.)
Map of 1.3 mm dust continuum emission covering about 2 pc
in the $\rho$ Ophiuchi cloud, with a resolution of 15\as\ (Motte et al 1998).
The emission is proportional to column density, and the lowest contour
corresponds to $N \sim 1.4\ee{22}$ \cmc\ or $\av \sim 14$ mag.
Contour levels increase in spacing at higher levels (see Motte et al 1998
for details). The highest contour implies $\av \sim 780$ mag.}
\end{figure}

\begin{figure}[ht!]

\caption{
(NOTE: this figure is not supplied here. It can be seen in the original paper
(ref. below) in grayscale.)
The 850 $\mu$m emission and 450 $\mu$m to 850 $\mu$m spectral index
distribution, covering about 7 pc with a resolution of 14\as\ at the
northern end of the Orion A (L1641) molecular cloud (Johnstone \& Bally 1999).
The Orion Nebula is located directly in front of the strong emission
at the center of the image.
({\it Left})
The 850 $\mu$m image showing the observed flux from $-0.1$ to 2 Jy/beam
with a linear transfer function. The highest flux level corresponds
roughly to $\av \sim 320$ mag if $\tk = 20$ K.
({\it Center})
The 850 $\mu$m image showing the observed flux from 100 mJy/beam
to 20 Jy/beam with a logarithmic transfer function.
({\it Right})
The 450 $\mu$m to 850 $\mu$m spectral index in the
range 2 to 6 with a linear transfer function.
}
\end{figure}

\end{document}